\documentclass[12pt]{iopart}
\usepackage{iopams}  
\expandafter\let\csname equation*\endcsname\relax

\expandafter\let\csname endequation*\endcsname\relax

\usepackage{amsmath}
\usepackage{graphicx}
\usepackage[usenames,dvipsnames,svgnames,table]{xcolor}
\usepackage{amsthm, amssymb, bbm, bm, mathtools}
\usepackage{enumerate}
\usepackage{graphicx}
\usepackage{float}
\usepackage{wrapfig, subcaption}
\usepackage[margin=0cm]{caption}
\usepackage[titletoc, toc]{appendix}
\usepackage{microtype}
\usepackage{tabularx}
\usepackage{booktabs,colortbl}
\usepackage{nicefrac}
\usepackage{blindtext}
\usepackage{setspace}
\usepackage{marginnote}
\usepackage{multirow}
\usepackage{csquotes}

\usepackage{mathptmx}

\usepackage[scr=boondoxo]{mathalfa}

\usepackage{microtype}

\usepackage{amsthm, amssymb, bbm, bm, mathtools}
\usepackage{etoolbox}
\def \b {\mathscr{b}}
\let\a=\alpha  \let\d=\delta
   
   \let\x=\xi

   \let\P=\Pi
   
  \let\th=\theta

\def\DD{{\cal D}}\def\AA{{\cal A}} \def\SS{{\cal S}}
\def\KK{{\cal K}}  
\def\ZZ{{\cal Z}}

\def\dde{\mathrm d}
\def\uwo{\underline w^{(0)}}

\def\to{\rightarrow}

\newcommand{\beq}{\begin{equation}} \newcommand{\eeq}{\end{equation}}

\begin{document}

\title[Stochasticity helps gradient-descent algorithms to navigate rough landscapes]{Stochasticity helps to navigate rough landscapes: comparing gradient-descent-based algorithms in the phase retrieval problem}

\author{Francesca Mignacco, Pierfrancesco Urbani}

\address{Université Paris-Saclay, CNRS, CEA,
Institut de physique théorique,  91191, Gif-sur-Yvette, France.}
\ead{francesca.mignacco@ipht.fr}
\author{Lenka Zdeborová}
\address{SPOC laboratory, Ecole Polytechnique Fédérale de Lausanne (EPFL), Switzerland.}
\vspace{10pt}
\begin{indented}
\item[]April 2021
\end{indented}

\begin{abstract}
In this paper we investigate how gradient-based algorithms such as
gradient descent, (multi-pass) stochastic gradient descent, its persistent variant, and the Langevin algorithm
navigate non-convex loss-landscapes and which of them is able to reach
the best generalization error at limited sample complexity. 
We consider the loss landscape of the high-dimensional phase retrieval
problem as a prototypical highly non-convex example. 
We observe that for phase retrieval the stochastic variants of gradient descent are
able to reach perfect generalization for regions of control parameters
where the gradient descent algorithm is not. 
We apply dynamical mean-field theory from
statistical physics to characterize analytically the full trajectories of these algorithms in
their continuous-time limit, with a warm start, and for large system sizes. 
We further unveil several intriguing properties of the landscape and
the algorithms such
as that the gradient descent can obtain better
generalization properties from less informed initializations. 
\end{abstract}

\maketitle

\section{Introduction}
Algorithms based on gradient-descent (GD) are the workhorses of many machine
learning applications involving the optimization of a high-dimensional
non-convex loss function. In particular, stochastic gradient descent
(SGD) has proved to be extremely efficient in navigating complex loss
landscapes. However, despite its practical success, the theoretical
understanding of the reasons behind the good generalization properties
of the algorithm remains sparse. Empirical evidence suggests that the
interplay between the optimization algorithm and the landscape is
crucial to achieve good performances. It has been shown, for instance, that the loss
landscape of state-of-the-art deep neural networks is far from simple:
adversarial initialization can trap SGD into global minima with poor
generalization \cite{LPA19}. Therefore, understanding the dynamics of
SGD is paramount in machine learning and optimization.

Investigating the dynamics of stochastic gradient descent and the role
of the stochasticity is consequently an active direction of research. While the
practical success of SGD compared to GD is rather generally accepted,
it is still far from clear what is really the key factor responsible
for this. Cases where the superiority of SGD with respect to GD
was shown theoretically are sparse, but see e.g. \cite{abbe2020polytime,haochen2020shape}. One hurdle that appears in
theoretical analysis is how to properly define the continuous limit of
SGD. In the limit of learning rate going to zero, SGD is considered to
lead to the gradient-flow limit, see e.g. \cite{cheng2020stochastic}, thus the difference with gradient
flow disappears. For learning rate kept finite, a line of works characterizes SGD as a discretization of a continuous-time Langevin-type
process \cite{li2017stochastic,jastrzkebski2017three,hu2017diffusion,cheng2020stochastic}. The
dependence of the variance of the noise on the current-weights and time is, however, not
given in a closed form in these works and thus difficult to analyze
explicitly. Another line of work challenged the central-limit-theorem
assumption of finite noise-variance behind these works by proposing the stochastic noise is
heavy-tail distributed \cite{simsekli2019tail}. 
In our work, we instead consider a variant of the stochastic gradient
descent called persistent-SGD, as recently defined in
\cite{francesca2020dynamical}. Persistent-SGD has a well define flow
limit $\eta\to 0$ and our analysis thus does not require other assumptions
about the nature of stochastic noise. 

Learning theory and computer science usually proceed in a manner that
makes minimalistic assumptions on the data ditribution. Statistical
physics usually takes a complementary way of understanding well
prototypical settings that capture the essesce of the question. This
is the path we take in this paper and compare the behaviour of
gradient descent-based algorithms on a prototypical choice of data and
learning model leading to high-dimensional and non-convex landscape. Specifically, we consider the problem of phase retrieval where the task consists in recovering an unknown signal from
a set of observations -- the absolute value of the signal's
projections onto measurement vectors. This problem appears in a series
of applications, including optics \cite{optics1,Millane:90}, acoustics \cite{acoustics}, and quantum mechanics
\cite{quantum}. We will consider the problem where the measurements
are i.i.d. Gaussian vectors and the number of measurements $M$ is only
constant times the dimensionality of the signal $N$, $\alpha=M/N$.
We consider the
high-dimensional limit where both the number of training samples and
the input dimensions go to infinity, at ratio $\alpha$ of
order one, typically $2-3$. 
In this work we view the phase retrieval as a prototypical example of
a simple single-layer neural network where the measurement vectors correspond to
the input samples, and the signal corresponds to
the teacher-network weights. The measurements then correspond to the output
labels. In the spirit of learning with neural networks we are
interested in the corresponding generalization error, i.e. the ability to
predict labels for previously unseen samples. We stress that it is
not the goal of this work to provide a competitive algorithm for the
phase retrieval. In the setting considered in this paper (i.e. iid
Gaussian inputs and teacher produced labels) it was conjectured that
the approximate message passing algorithm cannot be beaten in the
large size limit \cite{Barbier5451}. Instead, the main goal of this
paper is to study the performance of gradient-based algorithms and the loss landscape of the phase retrieval problem serves us as a
high-dimensional intrinsically non-convex prototype having multiple spurious minima and only one
solution (with a $\mathbb{Z}_2$ symmetry) leading to perfect
generalization error. 

We note that the landscape of phase retrieval problem is
somewhat different than the one of deep neural networks, that are highly
overparametrized and present entire regions of solutions with
zero training error and a good generalization. Consequences of this
difference and thus relevance of the present work for learning with
state-of-the art neural networks is left for future work. Instead the
present work investigates the performance of gradient-based
algorithms in an archetypal non-convex high-dimensional setting
providing a benchmark to assess the role played by stochasticity in
non-convex optimization problems in general. 

Our main contributions can be summarized as follows:
\begin{itemize}
\item We perform a series of numerical simulations in order to assess the generalization performance of the gradient
descent, multi-pass stochastic gradient descent, its persistent-SGD
version, and the Langevin algorithm as a function of the
  control parameters (mini-batch size, persistence time,
  temperature).  Our experimental findings reveal that in the considered problem stochasticity is beneficial for
  generalization. We also shed light on the qualitative difference
  between the sources of noise in the algorithms.
\item We investigate the
  role of the warm start and we find that gradient descent can be
  trapped very close to the signal, while perfect recovery can be
  reached starting from less informed initializations. 
\item We then apply dynamical mean-field theory (DMFT) from statistical
  physics to provide an analytic characterization of the full
  trajectory of the continuous limit of the GD, persistent-SGD and
  Langevin algorithms in the high-dimensional limit where the number
  of samples and dimension are both large, but their ratio $\alpha =
  \mathcal{O}(1)$, at times linear in the dimension. We use the theoretical curve as a baseline to show that the
  observed behavior is not due to finite-size or finite-learning-rate effects. 
\end{itemize}

\paragraph{Further related works ---} In this paper, we consider the
phase retrieval problems with Gaussian measurements and signal in the
high-dimensional limit. The loss landscape complexity of this problem was studied
using the Kac-Rice method in \cite{maillard2020landscape}, however,
bringing this analysis to concrete results seemed to be technically challenging. 
Signal recovery in this problem was studied
from the information-theoretic point of view and using approximate
message passing (AMP) algorithms that are considered optimal among all
polynomial algorithms for this case \cite{Barbier5451,mondelli2018fundamental,ma2019optimization}. In particular
it is known that while information-theoretically zero generalization
error can be reached for $\alpha>1$, the AMP algorithm is able to
do so for $\alpha>1.13$.

Performance of gradient descent for phase retrieval is worse than the
one of AMP in terms
of sample complexity and also
harder to analyze. In practice, one often uses gradient descent
initialized spectrally \cite{jonathan}, i.e. in the eigenvector corresponding to the
leading eigenvalue of a suitable matrix constructed from the labels and
the measurement vectors \cite{luo2019optimal}. Such spectral initialization is also
motivating our use of warm start that is mimicking it.  Concerning
randomly initialized gradient descent, \cite{Chen_2019} showed that
gradient descent needs a training set of size $\sim \mathcal{O}(N{\rm
  poly}(\log N))$ to retrieve the hidden signal. Several other works
in computer science consider gradient descepnt-type algorithms for
phase retrieval requiring  $\mathcal{O}(N{\rm
  poly}(\log N))$ samples \cite{ma2018implicit}.  
 The analysis carried
out in \cite{mannelli2020} suggests that the randomly-initialized algorithm can achieve
perfect generalization with much lower linear sample complexity. Authors of
\cite{cai2021solving} then show that linear (with unspecified large constant) sample complexity is achievable with
randomly initialized gradient descent for a suitably chosen loss
function. Finally \cite{mannelli2020optimization} have shown that over-parametrization can
bring the sample complexity of randomly initialized gradient descent
down to $\alpha=2$. While in the present work we will not be considering
overparametrization, we are interested in performance of
gradient-based algorithms for similarly small sample complexity
$\alpha$. We will be investigating several gradient-based algorithms
and judge their performance by the number of samples they require for
recovery of the signal. The fewer samples the better. This is why we
focus on the regime of $\alpha = \mathcal{O}(1)$.

The online SGD for phase retrieval has been studied, e.g., in \cite{tan2019online}.  Results for (multi-pass) stochastic gradient-descent in phase
retrieval are not known up to our best knowledge.
A theoretical understanding of the performance of (multi-pass) SGD at
small sample complexity requires
taking into account the full trajectory of the algorithm which is
challenging and done in the present paper.

\label{intro}
\section{The model and the data}
\label{sec:model_main}
We study the supervised learning problem of recovering an
$N-$dimensional real-valued vector
$\uwo=\{w_1^{(0)},\ldots,w_N^{(0)}\}$ from a set of $M=\alpha N$
real-valued noiseless measurements $\underline \xi^{\mu}=\{\x^\mu_1,\ldots, \xi^\mu_N\}$ of dimension $N$. We consider the signal $\uwo$ to be extracted with the uniform measure on the $N$-dimensional hyper sphere $|\uwo|^2=N$. We take the components of the vectors $\underline \xi^{\mu}$ to be  i.i.d. Gaussian random variables with zero mean and unit variance.
The non-linear measures of the signal vector $\uwo$ are  encoded in the labels
\beq
y^\mu=\biggr\rvert\frac{1}{\sqrt{N}} \underline \xi^\mu \cdot \underline w^{(0)}\biggr\rvert,\ \ \ \ \ \forall \mu=1,\ldots, M.
\eeq
We note that in applications the complex-valued phase retrieval is
more relevant, yet for the purpose of the present paper, which is
studying the performance of the gradient-based algorithms, the
real-valued version is sufficiently rich. 
We consider learning with a single-layer neural network by the minimization of the empirical risk
\beq
{\cal L}\left(\underline{w}\right) = \sum_{\mu=1}^{M} v(h_\mu; h^{(0)}_\mu),\label{loss}
\eeq
where $v$ is a cost function having a global minimum at $h_\mu=h^{(0)}_\mu$ and we have defined 
\beq
h_\mu = \frac 1{\sqrt N} \underline \xi^\mu \cdot \underline w, \ \ \ \ \ \ h_\mu^{(0)}= \frac 1{\sqrt N} \underline \xi^\mu \cdot \uwo.\label{gap}
\eeq
In what follows we consider a loss of the form: 
\beq
v(h,h_0)=\frac 14 (h^2-h_0^2)^2.
\eeq
However, the analytic derivation of the DMFT can be carried out for every twice-differentiable function $v$. Note that the empirical risk depends on the labels $y^\mu$ only through $h_\mu^{(0)}$. We consider a particular regularization of the weights where the training dynamics of $\underline{w}(t)$ is constrained on the hyper-sphere. In  \ref{sec:ridge}, we show that our results hold in a qualitative same manner for the more standard ridge regularization. 
\section{The analyzed algorithms}
\label{sec:analyzed_algorithms}
In this section we define the gradient-descent-based algorithms under consideration and their continuous-time limit that will then be studied using dynamical mean-field theory. 
The discrete dynamics of full-batch gradient descent is given by the weights update: 
\beq
\begin{split}
w_i(t+\eta)=w_i(t)-\eta\,\left[ \partial_{w_i}{\cal L}\left(\underline{w}\right)+\hat \nu(t) w_i(t)\right]\\=w_i(t)-\eta\left[\sum_{\mu=1}^{\a N} \partial_1 v(h_\mu; h^{(0)}_\mu) \frac 1{\sqrt N} \xi^\mu_i + \hat \nu(t) w_i(t)\right],\label{GDalgorithm}
\end{split}
\eeq
for all $i=1,...N$, where $\eta>0$ is the discrete time step and $\partial_1 v(h; h_0)$ indicates the derivative of the loss function with respect to its first argument. We have introduced a Lagrange multiplier $\hat \nu(t)$ to enforce the spherical constraint on the weights at each time step.  This is equivalent to a projection on the sphere at each iteration,  which is how we implement the numerical simulations. In the following, we analyze  different ways to add stochasticity to the dynamics. 
\paragraph{Multi-pass stochastic gradient descent dynamics ---}  
We study multi-pass stochastic gradient descent, where the samples are reused multiple times during training. The mini-batches are sampled with replacement with size $B=\b M$, $\b\in (0,1]$ at each time step. If we introduce a set of binary variables $s_{\mu}(t)\in\{0,1\}$, $\mu=1,...,M$ to select which samples are used compute the gradient, then in the large $N$ limit the vanilla-SGD algorithm described above is equivalent to draw
\beq
s_{\mu}(t)=\begin{cases}1 \,\,\,{\rm w.p. }\,\,\b\\
0 \,\,\,{\rm otherwise }
\end{cases}
\eeq
independently at each time step. However, the continuous-time limit
$\eta\rightarrow 0$ different from the gradient-flow is not well-defined in this case.
As done in \cite{francesca2020dynamical}, in order to consider the continuous-time dynamics, we define a discrete-time process for the variables $s_{\mu}(t)$ that has a well-defined continuous-time limit. We divide the time interval in finite steps of size $\eta$ and we define the following \emph{persistent} version of the stochastic gradient descent algorithm:
\beq
\begin{split}
s_{\mu}(t=0)=\begin{cases}1 \,\,\,{\rm w.p. }\,\,\b\\
0 \,\,\,{\rm otherwise }
\end{cases},\\
{\rm Prob}\left(s_{\mu}(t+\eta)=1|s_{\mu}(t)=0\right)=&\frac{\eta}{\tau},\\
{\rm Prob}\left(s_{\mu}(t+\eta)=0|s_{\mu}(t)=1\right)=&\frac{(1-\b)}{\b\tau}\eta.
\end{split}
\label{persistentSGD}
\eeq
In this case, each pattern stays out of the training mini-batch for a
typical time $\tau$, that we will refer to as the persistence time.
The stochastic gradient flow (SGF) dynamics is obtained by taking the $\eta\rightarrow 0$ limit of Eq. \eqref{persistentSGD}
\beq
\begin{split}
\frac{\partial w_i(t)}{\partial t} =
&-\frac 1\b\sum_{\mu=1}^{\a N} s_{\mu}(t)\partial_1 v(h_\mu(t);h_\mu^{(0)} )\frac{1}{\sqrt N} \xi^\mu_i\\&-\hat \nu(t) w_i(t) ,
\end{split}\label{SGF}
\eeq
for all $i=1,...N$, where we have rescaled the gradient by the
fraction of samples in the mini-batch. Note that, in this setting,
there are two parameters controlling the stochasticity of the
algorithm: the mini-batch size $\b$ and the persistence time
$\tau$. The standard SGD algorithm is recovered from (\ref{persistentSGD}) by setting $\tau=\eta/\b$ and finite learning rate $\eta$. In \ref{sec:numerical_simulations_supmat}, we show by numerical simulations with decreasing learning rate that the $\eta\rightarrow 0$ limit of the persistent SGD algorithm is different than gradient descent, while this is not the case for standard SGD.
\paragraph{Langevin dynamics ---}
A different kind of stochastic dynamics is provided by the Langevin algorithm at temperature~$T$, whose flow limit is defined by the following system of stochastic differential equations:
\beq
\begin{split}
\frac{\partial w_i(t)}{\partial t} = &-\sum_{\mu=1}^{\a N}\partial_1 v(h_\mu(t);h_\mu^{(0)} )\frac{1}{\sqrt N} \xi^\mu_i + \varsigma_i (t)\\&-\hat \nu(t) w_i(t).
\label{Langevin}
\end{split}
\eeq
for all $i=1,...N$.  The random vector $\underline{\varsigma} (t)$ is Gaussian white noise:
\beq
\begin{split}
\langle\varsigma_i(t)\rangle &=0,\qquad\qquad\qquad\quad \forall i=1,...N,\\
\langle \varsigma_i(t)\varsigma_j(t')\rangle &=2T\delta_{ij}\delta(t-t'),\quad \forall i,j=1,...N.
\end{split}
\eeq
Note that by setting $\b=1$ in Eq. \eqref{SGF} or $T=0$ in Eq. \eqref{Langevin} we recover the full-batch gradient-flow algorithm. 
\paragraph{Warm initialization ---}
In order to explore the energy landscape more thoroughly we consider here, next to the usual random initialization, informed/warm initializations.  
We initialize the weight vector as follows:
\beq 
\underline{w}(t=0)=m_0\, \underline{w}^{(0)}+c\,\underline{z} \in \mathbb{R}^N,\label{init}
\eeq
where $m_0>0$ is (on average) the initial projection of the weight vector onto the signal, i.e., the average \emph{magnetization} 
\beq
m(t)=\frac 1 N\underline w(t) \cdot \uwo\label{magnetization}
\eeq
at time $t=0$. The components of $\underline{z}$ are i.i.d. standard Gaussian variables and the coefficient $c$ is such that
$|\underline{w}(t=0)|^2=N$. Note that the warm initialization breaks
the $\mathbb{Z}_2$ symmetry of the problem. Therefore, in the
following $m(t)\in(0,1]$, $\forall t$. We stress here that while in
learning we are usually concerned with the test error/performance, in
the setting considered here (under the spherical constraint) the test
error is monotonic in the magnetization, see  \ref{sec:generalization_error}.  Thus,  in the following we directly use the magnetization
as a measure of accuracy. 
\\\\
This warm initialization can be thought of as a proxy for algorithms
where gradient-descent (or its variants) is run after the weights have
been spectrally initialized, i.e. using the principal eigenvalue of a
given pre-processing matrix as initial guess for the
weights. Spectrally initialized GD is used in a range of
applications, see e.g. \cite{jonathan}, as well as studied
theoretically, see e.g. \cite{mondelli2020optimal}. 
Warm initialization is formally needed to obtain non-trivial results for times linear in the dimension. Indeed, because of the $\mathbb{Z}_2$ symmetry we would need time at least logarithmic in dimension in order to escape the space of weights uncorrelated with the solution. This was referred to as the ``escape from mediocrity'' in \cite{ben2020classification}. 

\section{Characterization of the dynamics}\label{sec:analytic_characterization}
In this section we provide a closed-form characterization of the flow dynamics of the persistent-SGD and Langevin algorithms presented above in the high-dimensional limit. To this end, we apply dynamical mean-field theory from statistical physics \cite{MPV90,GKKR96,PUZ20}. This analytic framework is useful to study the stochastic evolution of large systems of interacting degrees of freedom \cite{SZipp81,CK93,SCS88}. DMFT has been rigorously proven in some specific cases \cite{arous1997symmetric}, but not yet in the present one. 
Here we present the main analytic results, more details are provided in  \ref{sec:derivationDMFT}. The derivation follows the line of \cite{ABUZ18,francesca2020dynamical}, for a different data structure and loss function. We also need to include the spherical constraint and Langevin noise in the dynamics. We consider the high-dimensional limit $N\rightarrow\infty$, at fixed sample complexity $\alpha=M/N$, mini-batch fraction $\b$, persistent time $\tau$ and temperature $T$. For simplicity, we regroup the flow dynamics of multi-pass SGD \eqref{SGF} and Langevin \eqref{Langevin} in the same equation:
\beq
\begin{split}
\frac{\partial w_i(t)}{\partial t} 
=-\hat \nu(t) w_i(t) + \varsigma_i (t)\\-\frac 1\b\sum_{\mu=1}^{\a N} s_{\mu}(t)\partial_1 v(h_\mu(t);h_\mu^{(0)} )\frac{1}{\sqrt N} \xi^\mu_i.\label{continuous_equation}
\end{split}
\eeq
The performance of the algorithms as a function of training time  is encoded in the magnetization $m(t)$ defined in Eq.~\eqref{magnetization}, that is equal to $1$ for perfect recovery of the signal. In the high-dimensional limit, we obtain that the evolution of the magnetization is described by the following deterministic differential equation:
\beq
\partial_tm(t) = -\hat \nu(t) m(t) - \mu(t),\ \ \ \ \ \ \ \ \ m(0)=m_0\:,
\eeq
where 
\beq
\begin{split}
\hat\nu(t)&=-\frac \alpha \b \langle \tilde h(t)s(t)\partial_1v(\tilde h(t);h_0))\rangle+T ,\\
\mu(t)&=\frac \alpha \b \langle s(t) h_0 \partial_1 v(\tilde h(t);h_0)\rangle,\\
\tilde h(t) &\equiv h(t) + h_0 m(t).
\end{split}
\eeq
The brackets $\langle \cdot \rangle$ stand for the average over different sources of noise: 
\begin{itemize}
\item the binary variable $s(t)$, distributed as in Eq. \eqref{persistentSGD} for $\eta\rightarrow 0$;
\item the standard Gaussian variable $h_0\sim \mathcal{N}(0,1)$; 
\item the effective stochastic process for the typical gap $h(t)$ defined in Eq. \eqref{gap}.
\end{itemize}
The evolution of $h(t)$ is given by
\beq
\begin{split}
\partial_t h(t) = &-\left( \hat \nu(t)+\delta \nu(t) \right)h(t) - \frac 1\b s(t)\partial_1v(\tilde h(t);h_0) \\&+ \int_0^t \dde t'\, M_R(t,t') h(t') + \chi (t),\label{effective_process}
\end{split}
\eeq
with initial condition
\beq
P(h(0))=\frac{1}{\sqrt{2\pi(1-m_0^2)}} e^{-h(0)^2/2(1-m_0^2)}.
\eeq 
The dynamical noise $\chi(t)$ in Eq. \eqref{effective_process} is Gaussian distributed, with 
\beq
\begin{split}
\langle \chi(t) \rangle &= 0, \\
\langle \chi(t) \chi(t')\rangle &= 2T\delta(t-t') +M_C(t,t').
\end{split}
\eeq The expressions for the kernels $M_C(t,t')$ and $M_R(t,t')$ and the auxiliary function $\delta{\nu}(t)$ are given by
\beq
\begin{split}
M_C(t,t') &=\frac{\a}{\b^2} \langle s(t)s(t') \partial_1 v(\tilde h(t);h_0)\partial_1 v(\tilde h(t');h_0)) \rangle,\\
M_R(t,t')&=\frac{\alpha}{\b^2} \frac{\delta}{\delta P(t')}\langle s(t)\partial_1  v(\tilde h(t);h_0)\rangle\biggr\rvert_{P=0},\\
\delta\nu(t)&=\frac{\alpha}{\b} \langle s(t) \partial_1^2v(\tilde h(t);h_0)\rangle,
\end{split}
\eeq
where $P(t')$ is a linear perturbation applied to $h$ at time~$t'$ and then set to zero, that we only need to define $M_R(t,t')$.
Overall, we obtain that the performance of the algorithms over time is described by a system of integro-differential equations that must be solved numerically in a self-consistent way. As done in \cite{francesca2020dynamical}, we start from a simple guess of the kernels and the auxiliary functions and we compute many realizations of the curve $h(t)$ in Eq. \eqref{effective_process}. We use these curves to update the kernels and we iterate the procedure until convergence. The details are relegated to  \ref{sec:numericsDMFT}. A first implementation of this procedure was proposed in \cite{EO92,EO94}. 
Once the stochastic process defined in Eq. \eqref{effective_process} has reached convergence, we can compute other quantities of interest. For instance, we can track the evolution of the average training loss defined in Eq. \eqref{loss} in the high-dimensional limit:
\beq
\ell(t)=  \langle  v(\tilde h(t);h_0)\rangle.
\eeq
These equations provide a dimension-independent way to track the
performance of the algorithm in the limit of
high-dimensions and infinitesimal learning rate as a function of time. Indeed, since the solution of the problem is planted and the measurements are noiseless, in this case zero training loss corresponds to zero generalization error. 
\\
Note that this formalism allows to study the dynamics of the
corresponding algorithms without any approximation on the distribution
of the noise introduced by stochasticity. This is at variance with the
works that consider SGD as a noisy approximation of GD invoking variants of
central limit theorem \cite{li2017stochastic,jastrzkebski2017three,hu2017diffusion,cheng2020stochastic,simsekli2019tail}. 
\begin{figure*}[!ht]
\begin{center}
\includegraphics[scale=0.181]{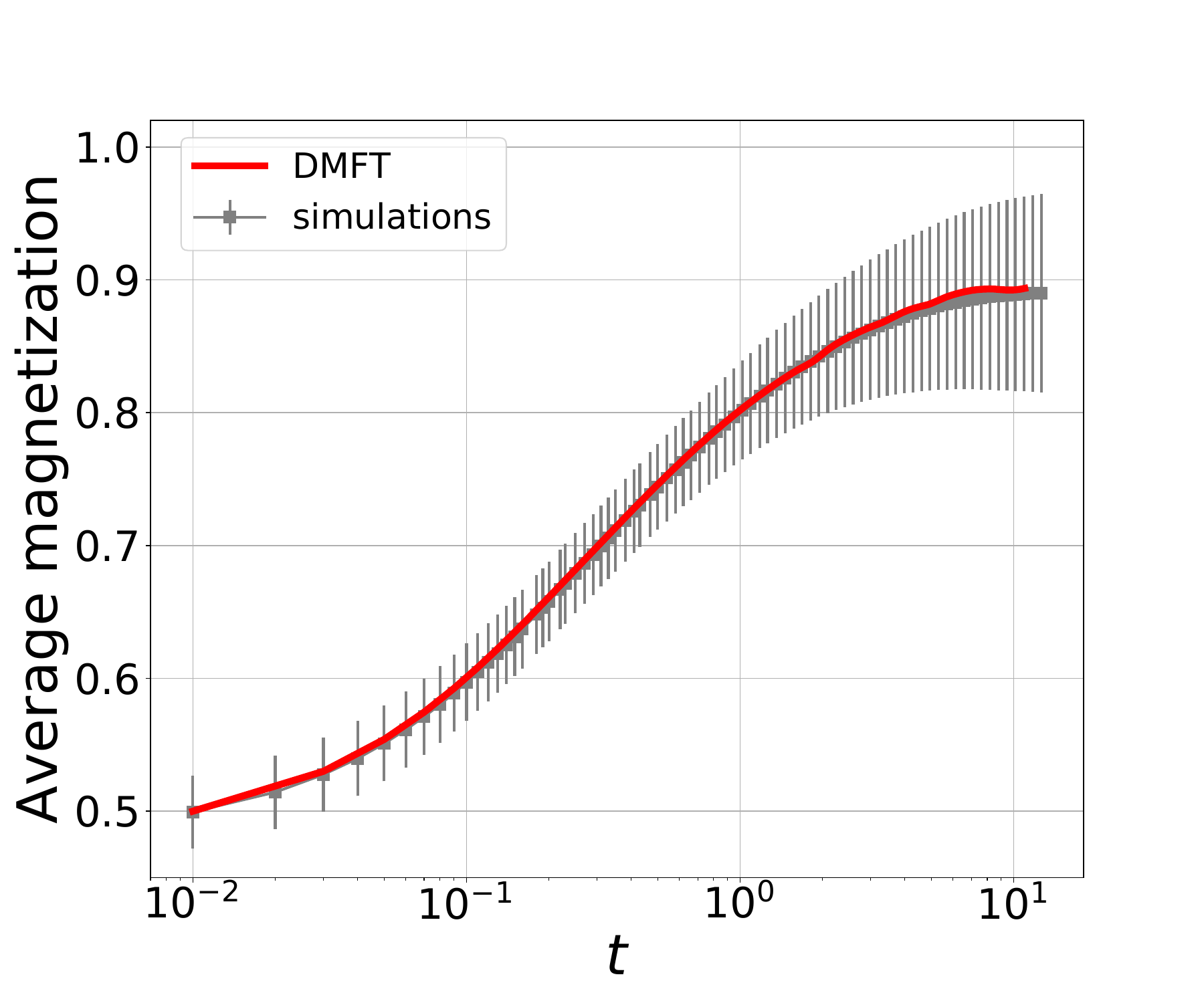}
\hspace{-7mm}
\includegraphics[scale=0.181]{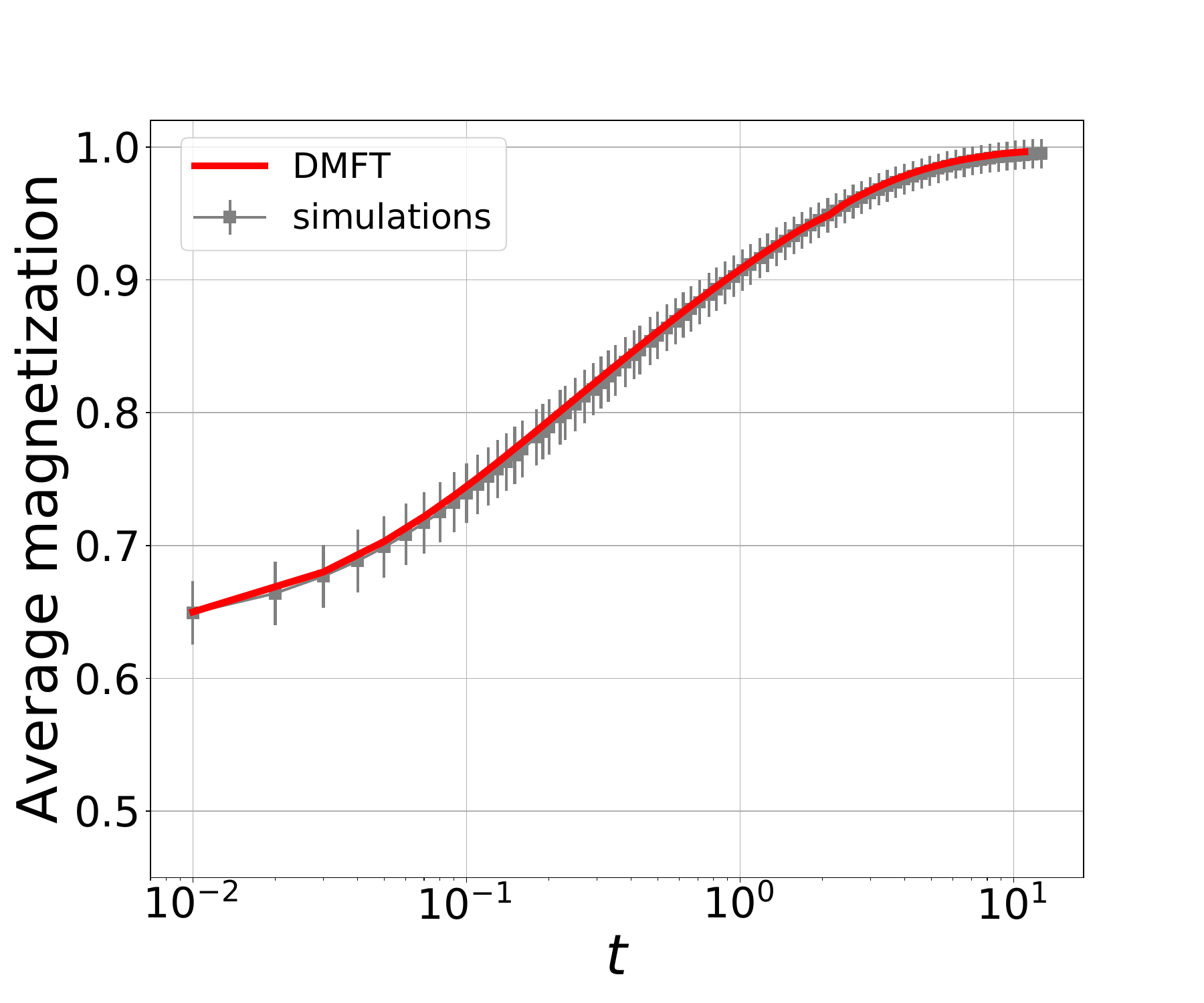}
\hspace{-7mm}
\includegraphics[scale=0.181]{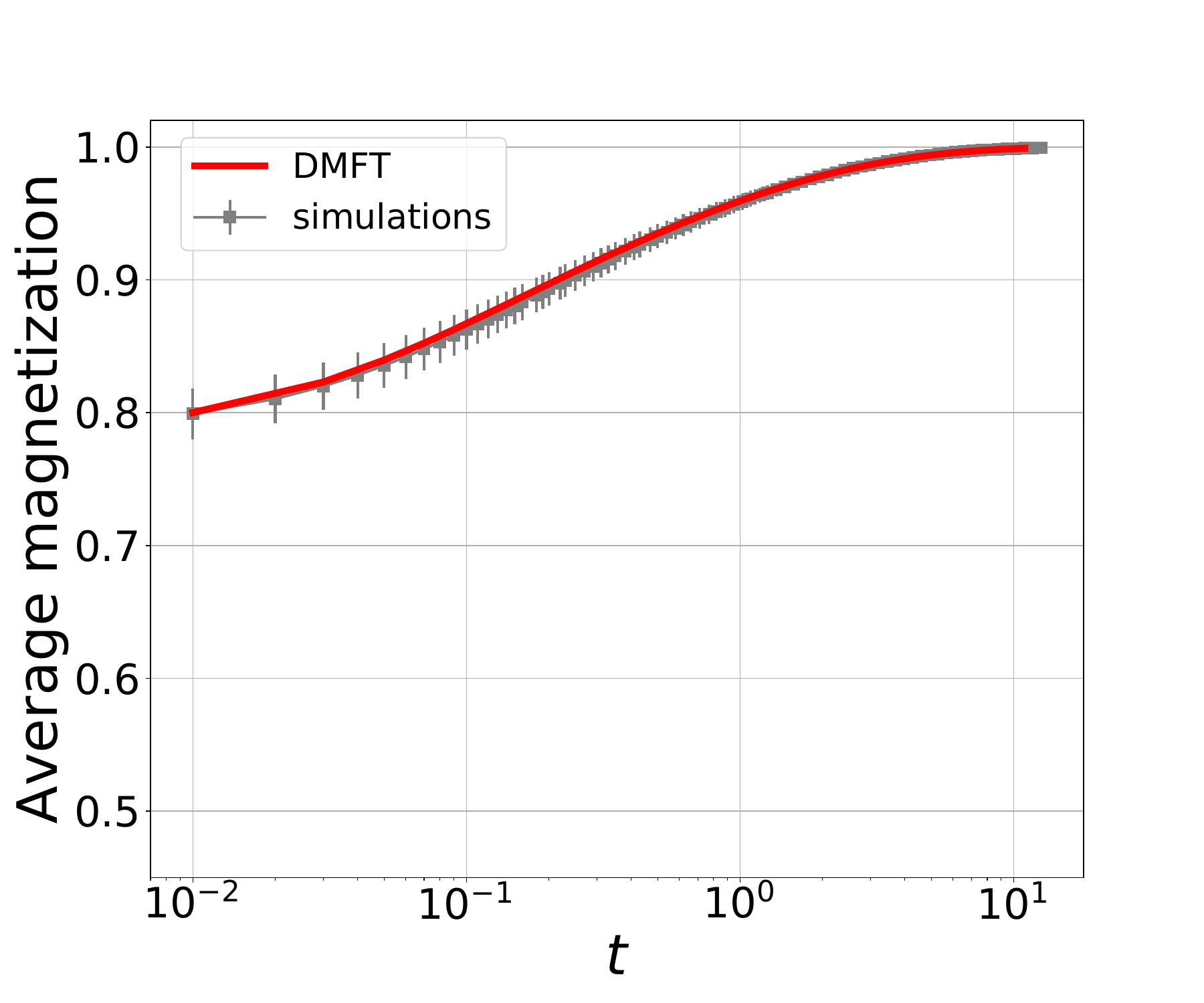}
\end{center}
\begin{center}
\includegraphics[scale=0.181]{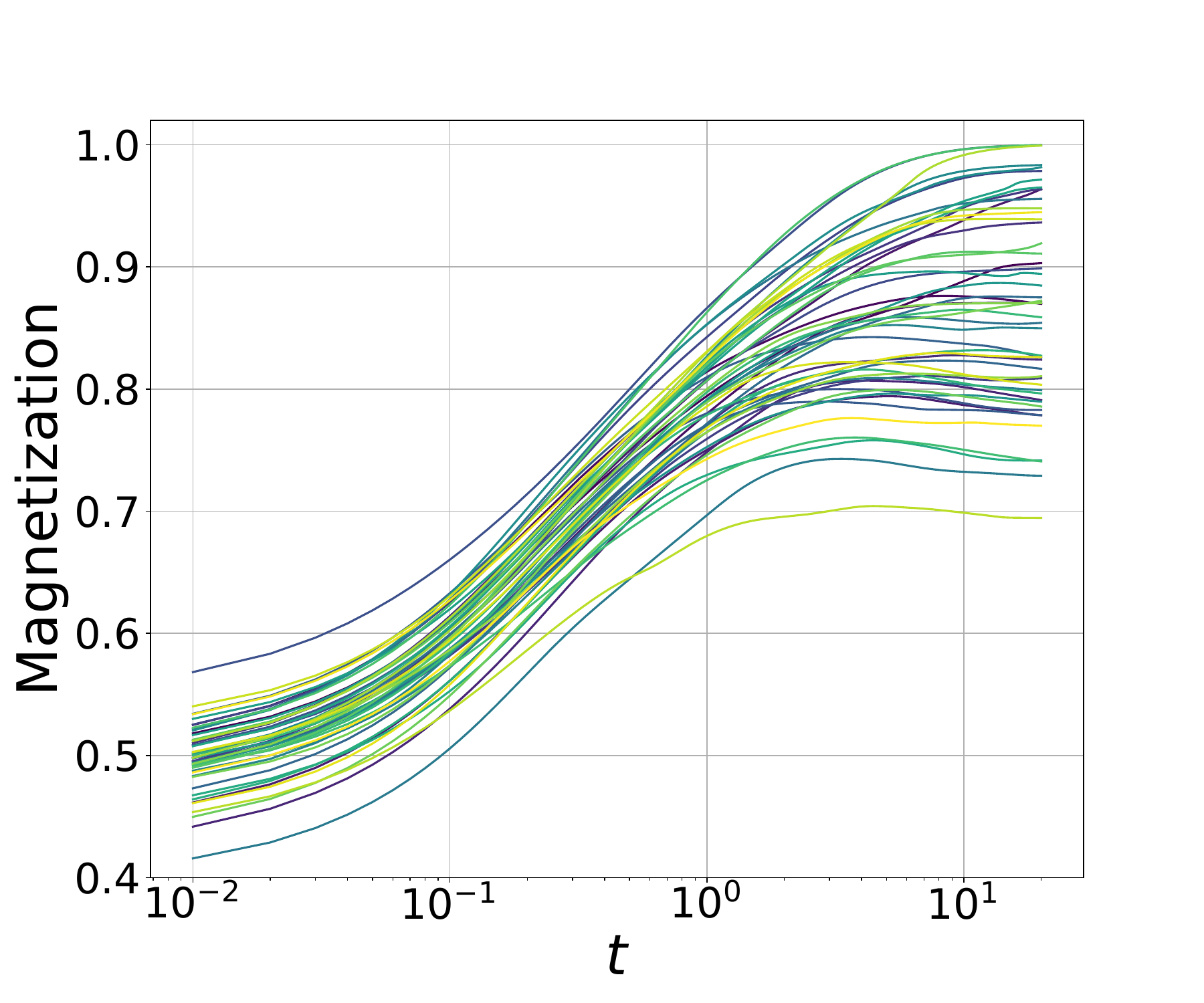}
\hspace{-7mm}
\includegraphics[scale=0.181]{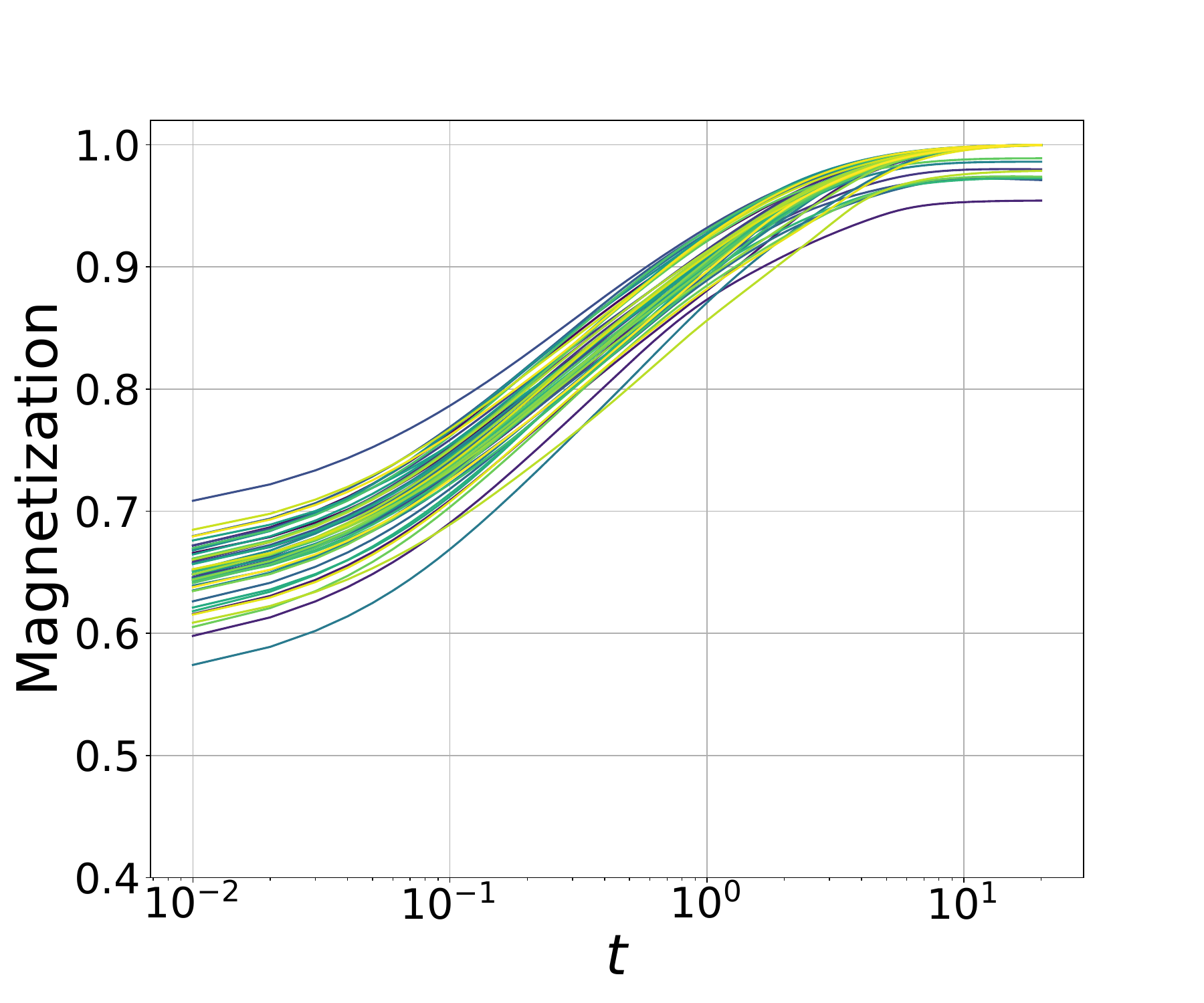}
\hspace{-7mm}
\includegraphics[scale=0.181]{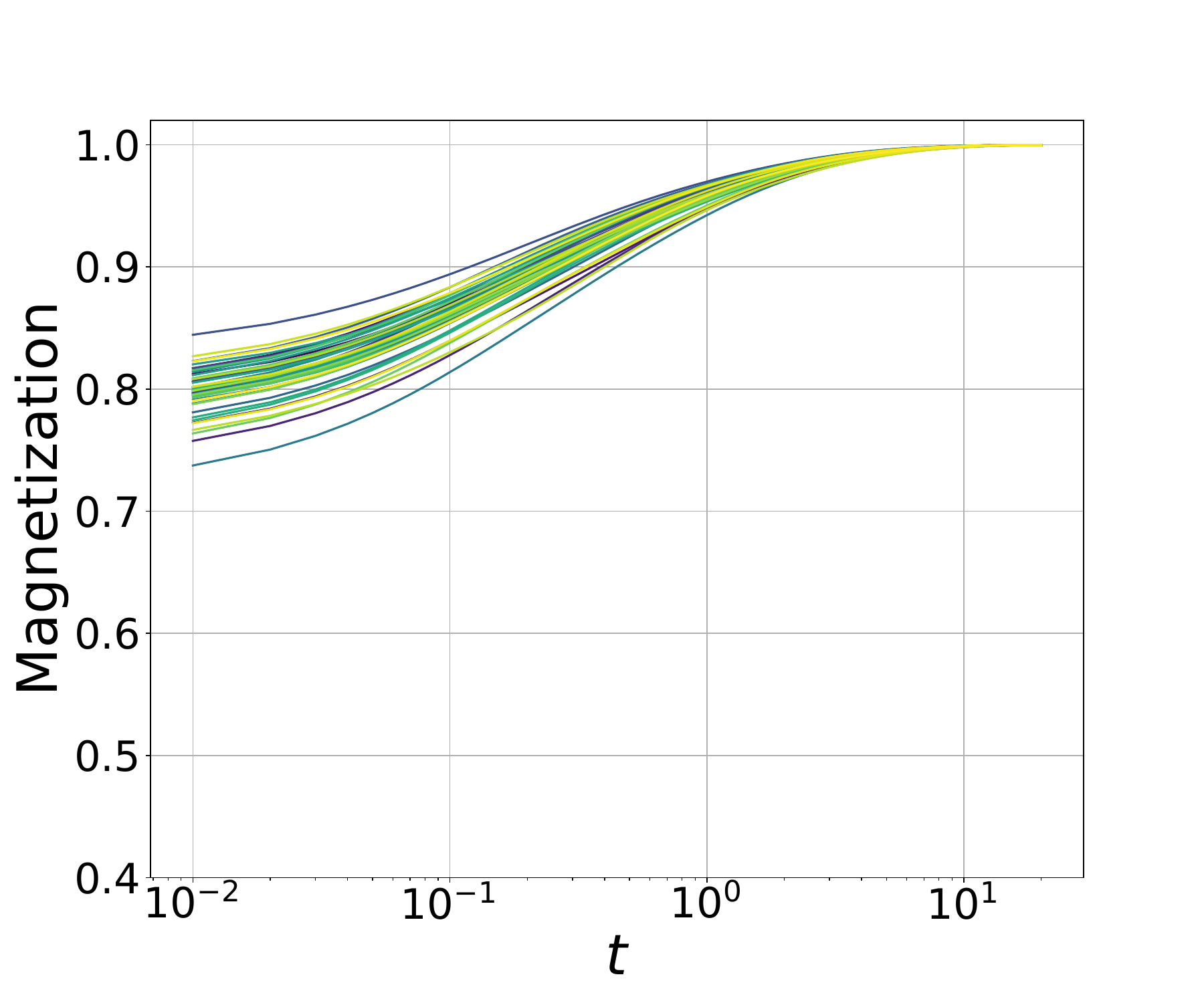}
\end{center}
\caption{\label{instances_m} {\bf Upper.} Average magnetization as a function of training time for the full-batch gradient descent/flow algorithm. We consider
  $\alpha=M/N=2$ and three different initializations: $m_0=0.5$ (left),
  $m_0=0.65$ (center), $m_0=0.8$ (right). The grey dots
represent numerical simulations  ($N=1000$, $\eta=0.01$),
  averaged over $1000$ seeds (generating a new dataset and signal for each seed). The full red line marks the theoretical prediction from DMFT
  obtained in the high-dimensional limit of gradient-flow. {\bf
    Bottom.}  We fix the landscape by fixing the dataset and we show $50$ instances of the magnetization as a function of time, for different realizations of the noise vector $\underline{z}$ defined in Eq.~\eqref{init}, at $m_0=0.5$ (left), $m_0=0.65$ (center), $m_0=0.8$ (right). For visibility purposes, we plot $t+\eta$ on the x-axes. }
\end{figure*}
\begin{figure}[!ht]
\begin{center}
\includegraphics[scale=0.266]{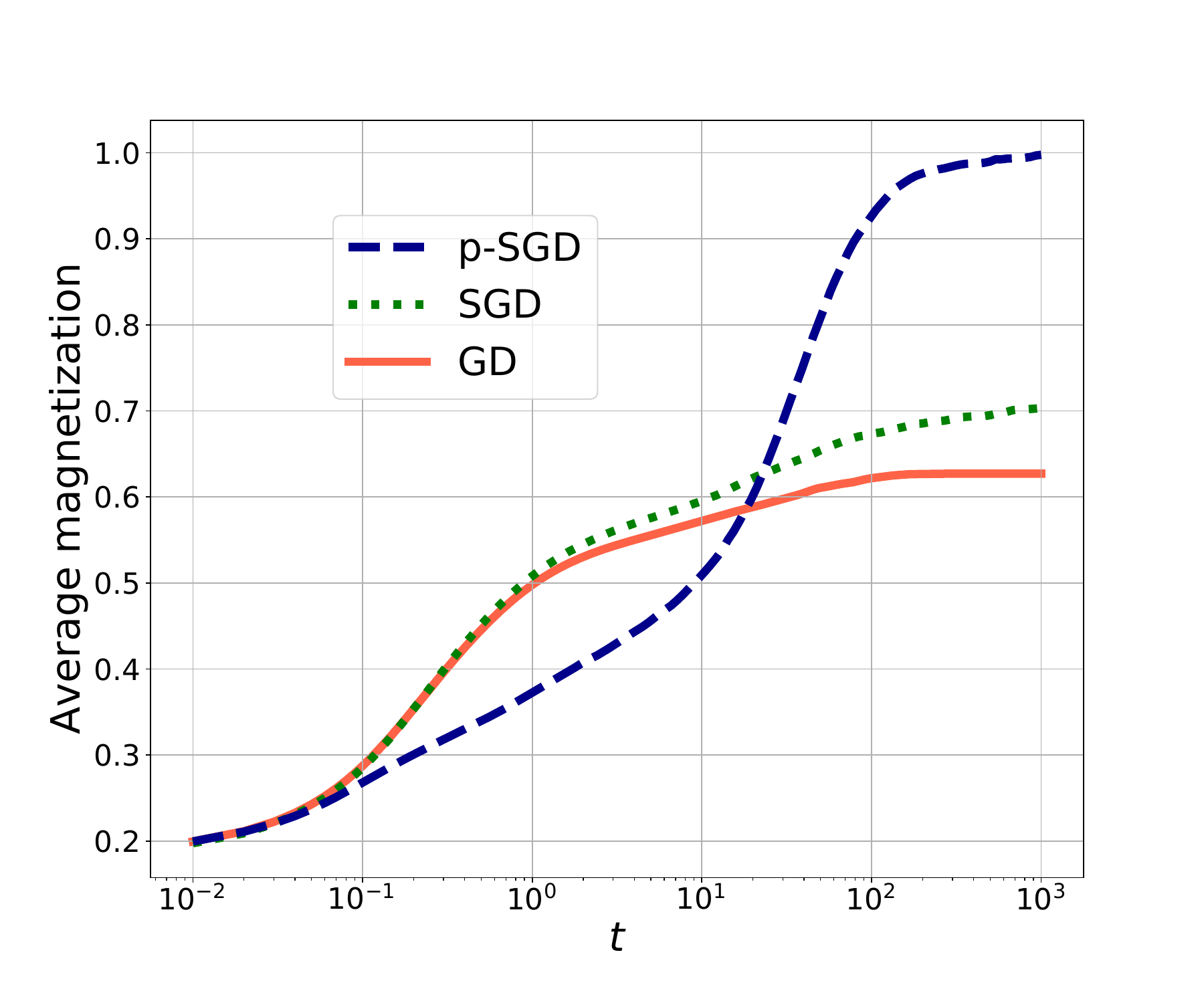}
\hspace{-7mm}\includegraphics[scale=0.266]{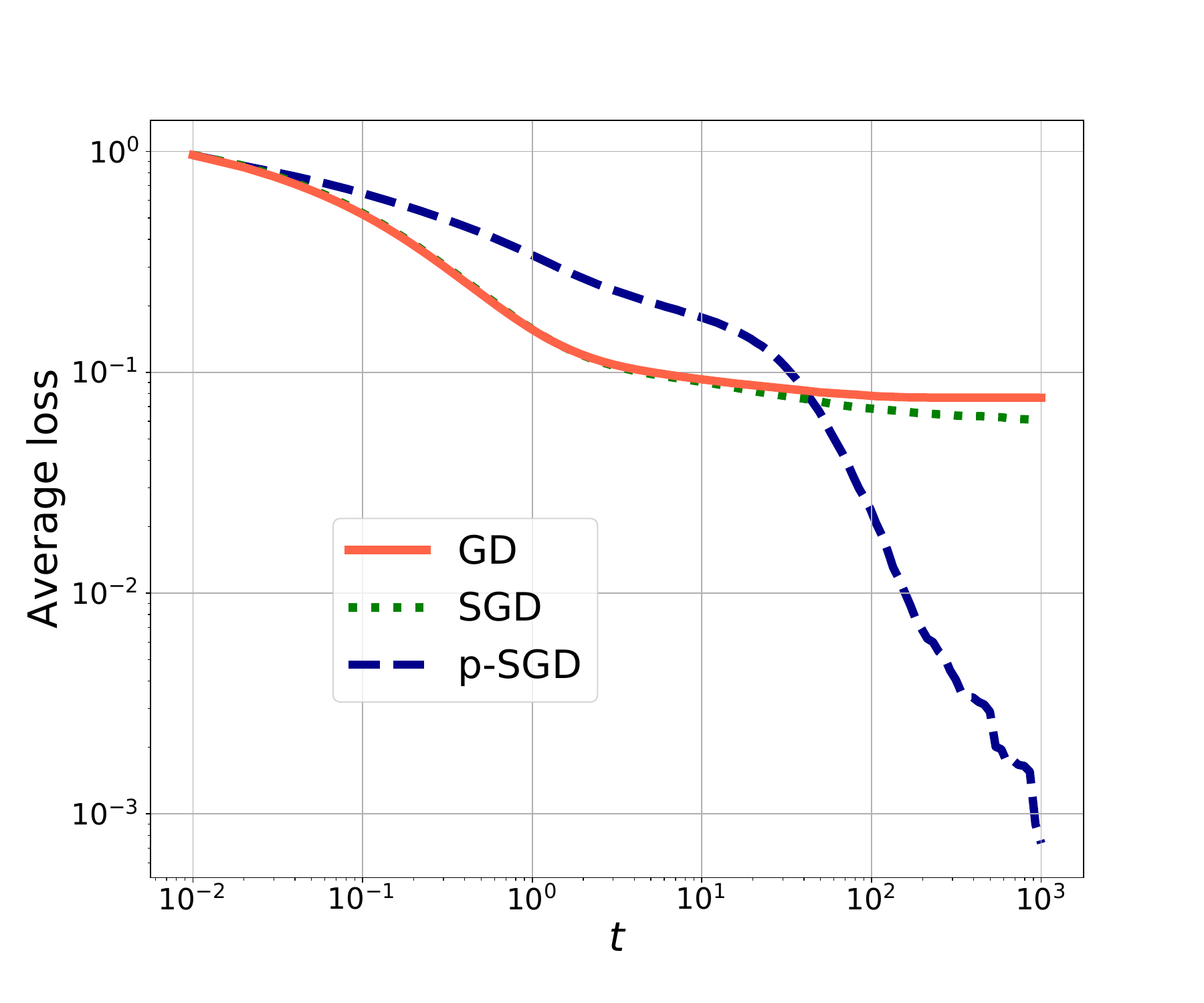}
\end{center}
\caption{\label{averageSGDvsGD} Average magnetization (left) and average training loss (right) as a function of training time, at fixed $\alpha=M/N=3$, initial magnetization $m_0=0.2$, input dimension $N=1000$, learning rate $\eta=0.01$. We show the performance of full-batch gradient descent (red line), multi-pass vanilla SGD at $\b=0.5$ (dotted green line), and persistent SGD at $\tau=1$, $\b=0.5$ (dashed blue line). The averages are computed over $500$ seeds (generating a new instance for each seed). At time $t=1000$, the percentages of seeds that have reached training loss below $10^{-7}$ are: $9\%$ (GD), $30\%$ (SGD), $99\%$ (persistent-SGD). For visibility purposes, we plot $t+\eta$ on the x-axes. }
\end{figure}

\begin{figure*}[!ht]
\centering
\includegraphics[scale=0.181]{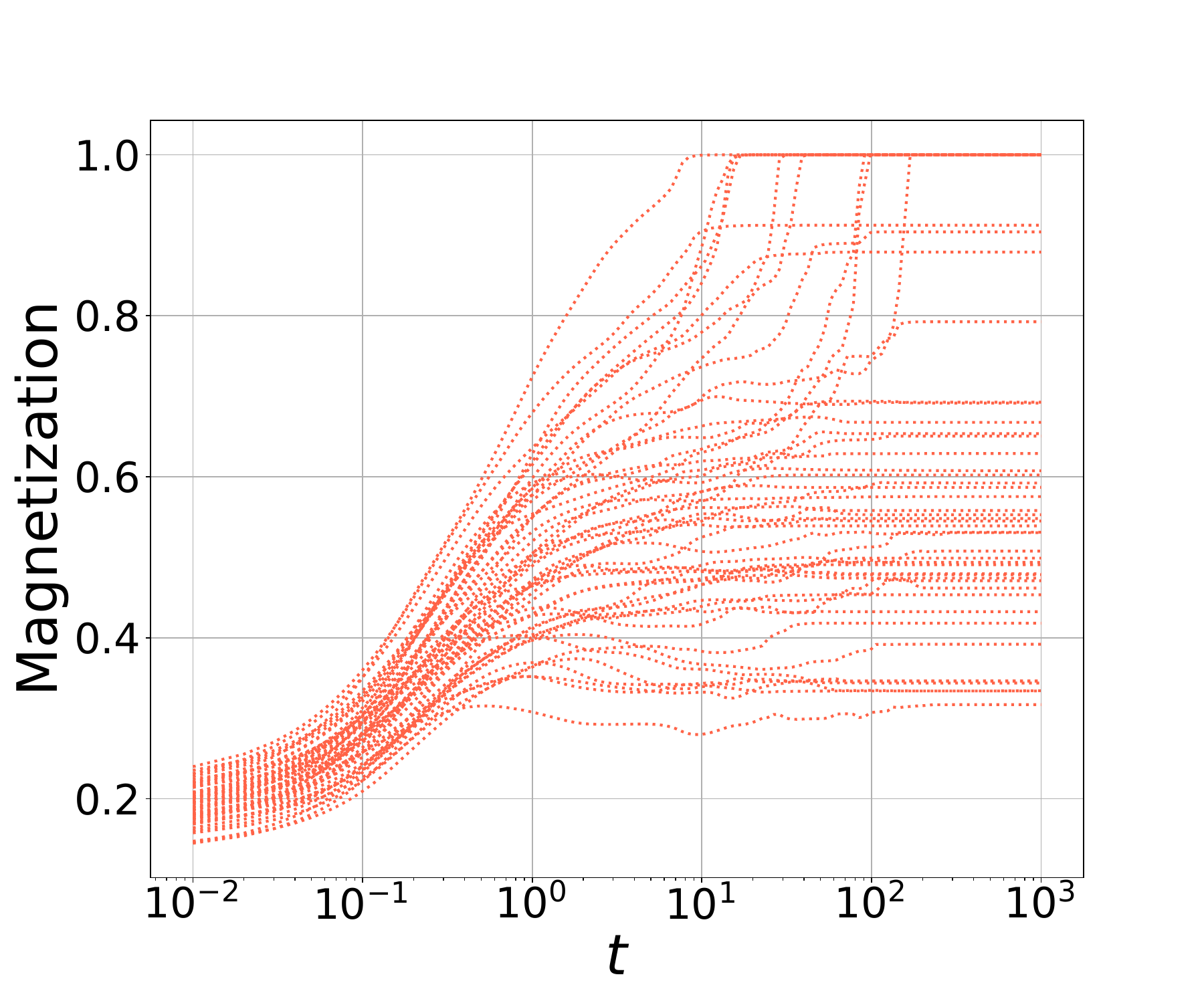}
\hspace{-7mm}
\includegraphics[scale=0.181]{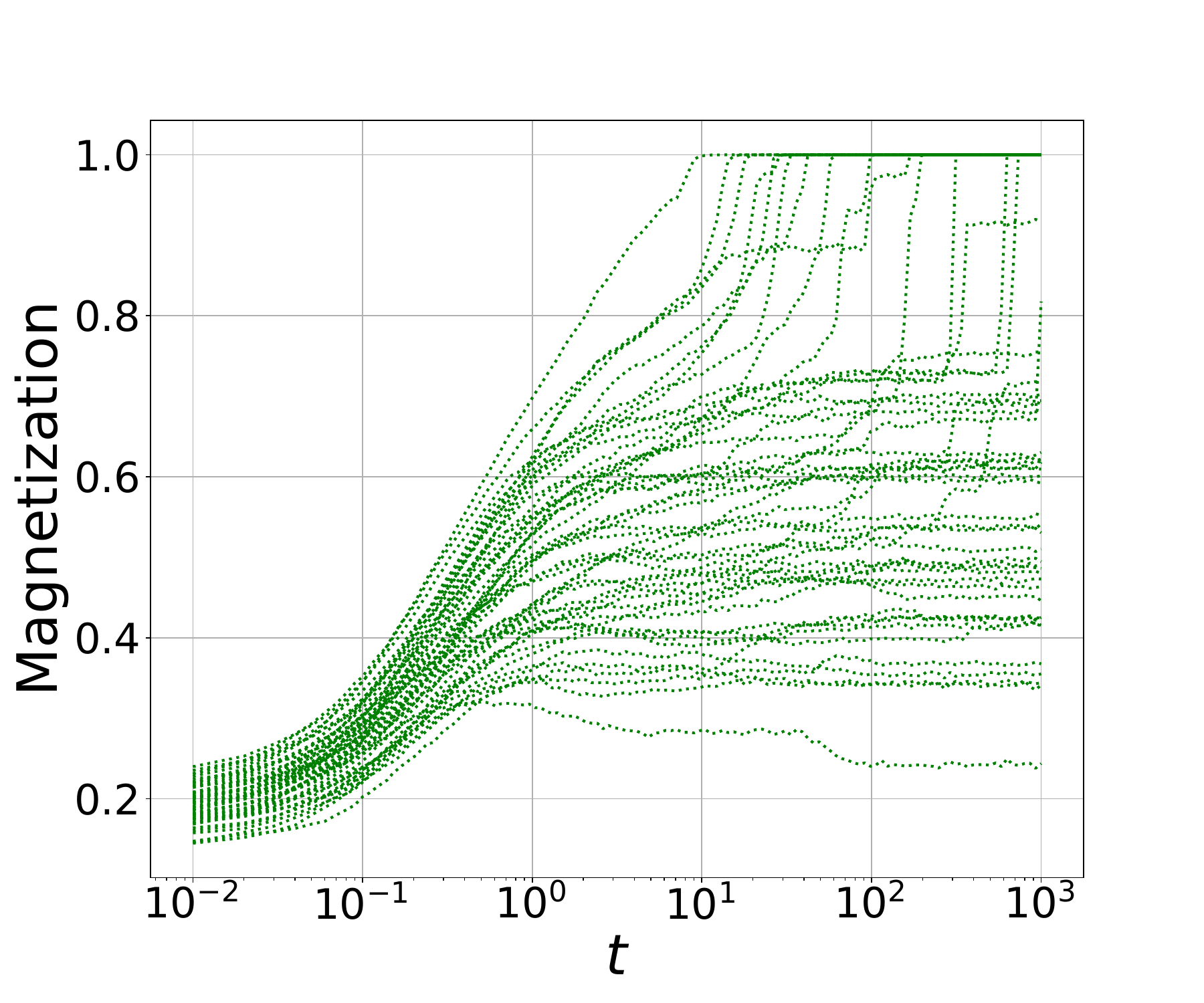}
\hspace{-7mm}
\includegraphics[scale=0.181]{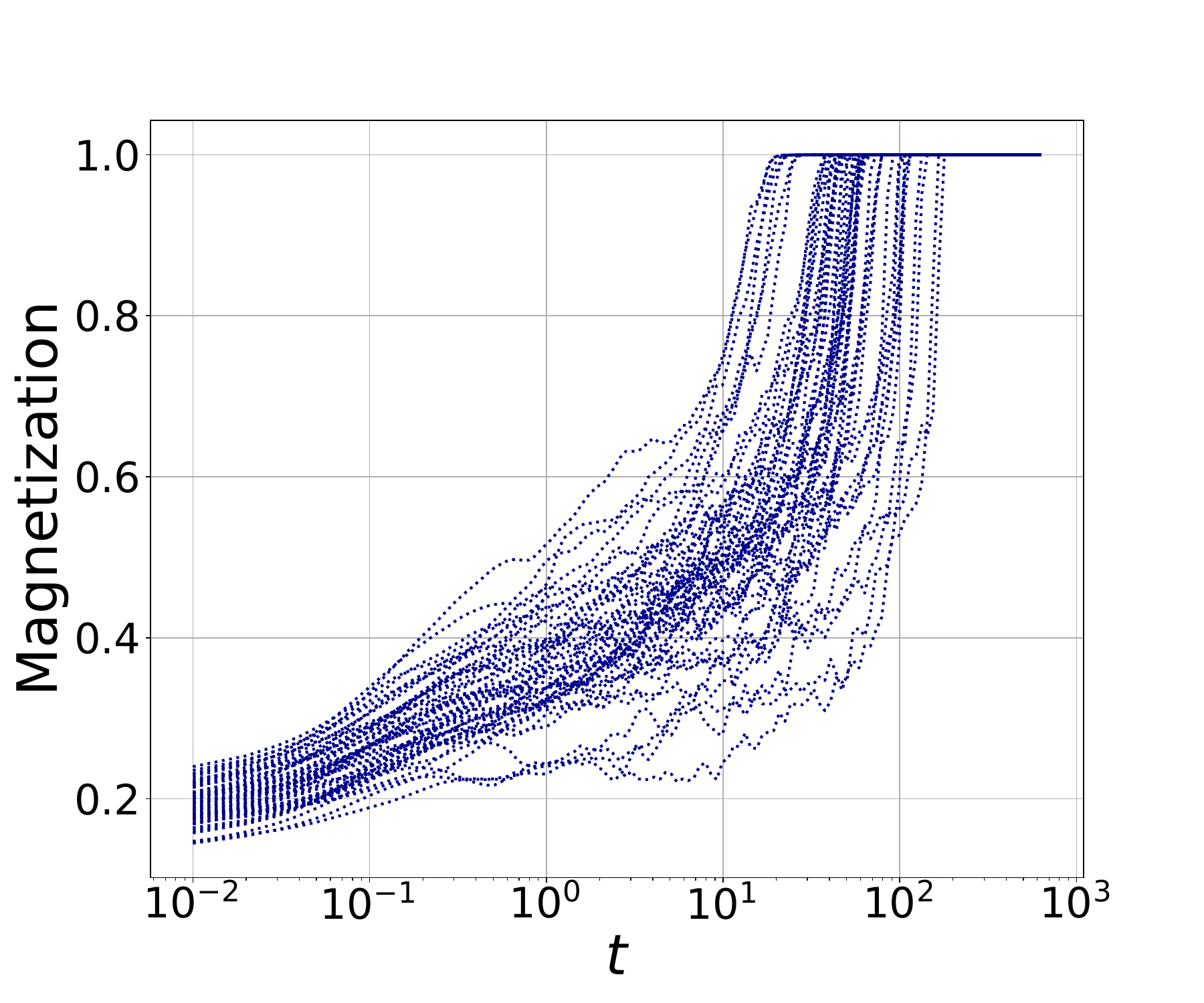}
\caption{\label{SGDvsGD}
We show $50$ instances of the magnetization as a function of training
time for GD (left), SGD at $\b=0.5$ (center) and persistent-SGD at
$\b=0.5$, $\tau=1$ (right). All other parameters are as in
Fig.~\ref{averageSGDvsGD}. For each seed, a new instance is generated. For visibility purposes, we plot $t+\eta$ on the x-axes.}
\end{figure*}

\section{Results for the dynamics}
\label{results_dynamics_main}
In this section, we discuss our findings on the dynamics of the
gradient-based algorithms under consideration. We compare the results from
simulations to the DMFT theoretical prediction. The DMFT is valid in
the continuous flow limit and in the high-dimensional limit. The
simulations are performed for sizes large enough and learning rate small enough
so that this limit is closely approached. This analysis sheds light on
how stochasticity helps to navigate the loss landscape and on the
impact of the different control parameters, notably the batch size $\b$,
temperature $T$, and persistence time $\tau$, on the test performance.
\paragraph{The trapping landscape ---} Fig. \ref{instances_m}
illustrates the performance of gradient descent starting from three
increasing initializations: $m_0=0.5$ (left), $m_0=0.65$ (center), and
$m_0=0.8$ (right) at $\alpha=2$, i.e. number of samples twice the
dimension. In the three lower panels, we plot the magnetization for different seeds --
corresponding to different realizations of the noise vector
$\underline{z}$ defined in Eq. \eqref{init} -- with a dataset, i.e. the inputs and labels, drawn at random
and fixed. The
evolution of different instances from simulations is thus probing
the very same loss-landscape, the figure then highlights the
complexity of the landscape. First, we observe that a warm start is
not enough to reach perfect recovery. This suggests that the landscape
is very rough, with multiple local minima at all heights. Indeed, we
see that gradient descent can get stuck even very close to the global minimum at $m=1$. 
From the right panel of the figure, we see that at time $t\sim 10$ all
seeds initialized with magnetization $m_0=0.8$ have achieved
perfect recovery $m=1$. 
However, the left and center panels show that some seeds starting at $m_0<0.8$ and
reaching  $m=0.8$ only at $t>0$ can get stuck for long times. 
Hence we deduce that the topological complexity of the landscape is
such that some regions of the weights space can be trapping even if
they are closer to the signal than others that do not trap the dynamics. We observe that a more informed initialization does not guarantee a
better generalization. 
This can be further seen comparing the left panel to the central
one. Indeed, we find that some seeds initialized at $m_0>0.6$ are
stuck at $m<1$ at time $t\sim 10$, while some seeds starting at
$m_0<0.6$ have already reached perfect generalization. 
Consequently, in this regime of parameters, the full trajectory of the algorithm is crucial to achieve perfect recovery. 
\\\\
In the upper panels of Fig. \ref{instances_m}, we compare the average
magnetization from numerical simulations at finite system size and finite learning rate (grey dots) to
the theoretical prediction (red line) obtained by integrating the DMFT
equations derived in the high-dimensional continuous limit. 
 In this case, we generate a new dataset for each
simulations in order to remove sample-to-sample fluctuations. 
We find a very good agreement
between asymptotic theory and the average from simulations already for
the used system sizes and learning rates, indicating that the observed
behavior is not a feature of finite size or finite learning rate
effects. Additional simulations supporting this evidence are left to
the  \ref{sec:numerical_simulations_supmat}. 

\paragraph{Multi-pass SGD outperforms GD ---} Fig.~\ref{averageSGDvsGD} shows
the average magnetization -- defined in Eq.~\eqref{magnetization} -- and the
average training loss -- defined in Eq. \eqref{loss} -- as a function of time
for full-batch gradient descent, multi-pass SGD and its persistent version.  In the case of multi-pass SGD, we sample (with replacement) minibatches of size $\b M$ at each time step. In Fig. \ref{SGDvsGD}, we depict 
different instances of the dynamics, corresponding to different
realizations of the dataset and the noise vector $\underline{z}$
(Eq. \eqref{init}). We find that SGD and persistent-SGD with $\tau=1$ outperform GD in recovering the
hidden signal. Indeed, at time scales at which persistent-SGD has already reached
magnetization one and zero loss, gradient descent is stuck in regions
of poorer generalization. The average magnetization of SGD lies between the two. Therefore, a finite batch size is
beneficial for the performance. Furthermore, the behavior of the
curves for different seeds unveils an important role played by the
persistence time. Indeed, while the evolution of the magnetization for
GD is characterized by long plateaus alternated by sudden jumps,
persistent-SGD is not stuck in the same region for long
times. Again, the behavior of SGD is intermediate between the two: we see from the central panel of Fig. \ref{SGDvsGD} that the disappearance of the plateaus is a feature of a finite persistence time.
These findings suggest that the interplay of the finite batch size and
the persistence time is crucial to achieve the optimal
performance. Additional simulations supporting this numerical evidence
are provided in  \ref{sec:numerical_simulations_supmat}.

\begin{figure*}[!ht]
\begin{center}
\includegraphics[scale=0.18]{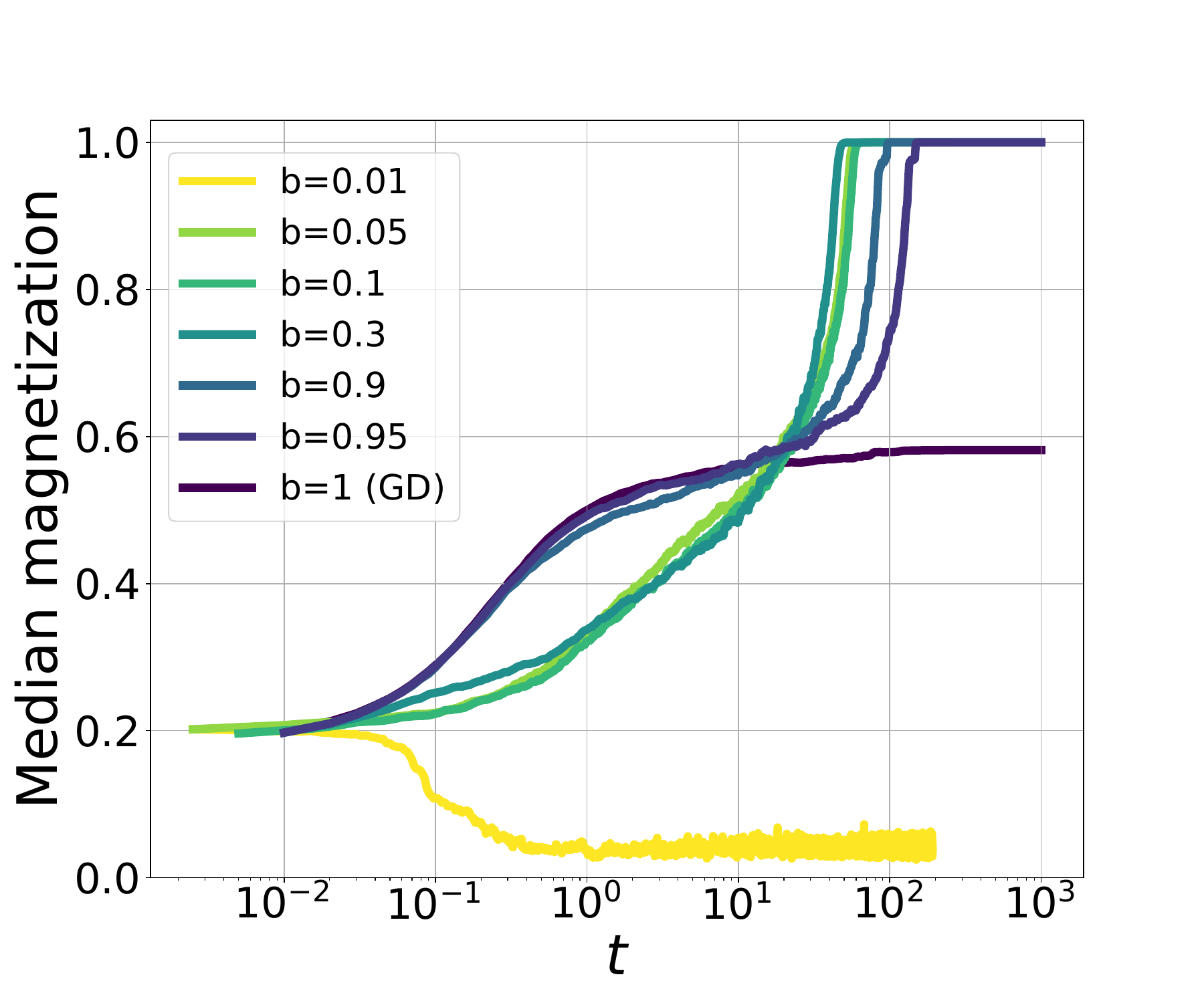}
\hspace{-7mm}
\includegraphics[scale=0.18]{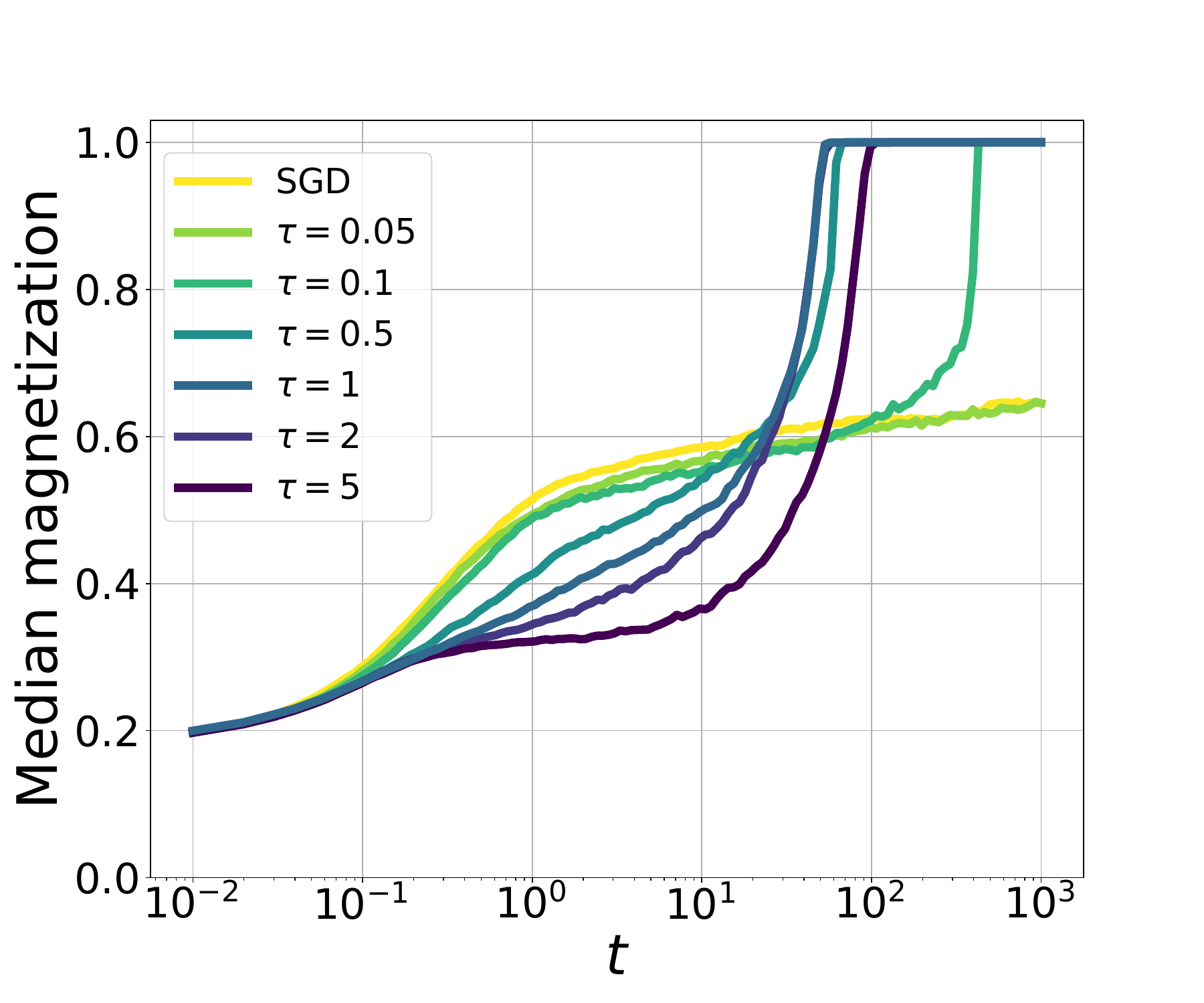}
\hspace{-7mm}
\includegraphics[scale=0.18]{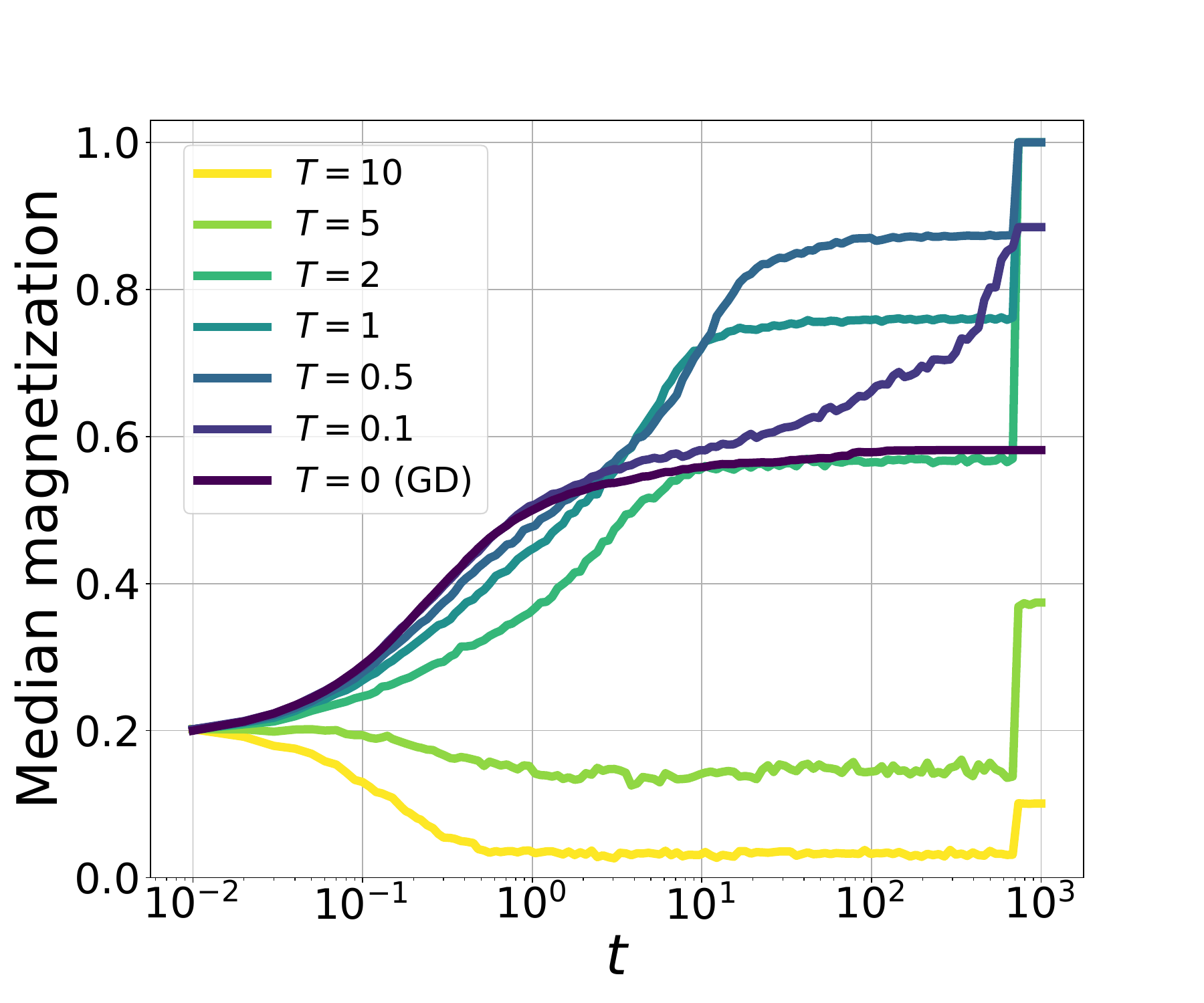}\\
\hspace{-6mm}
\includegraphics[scale=0.172]{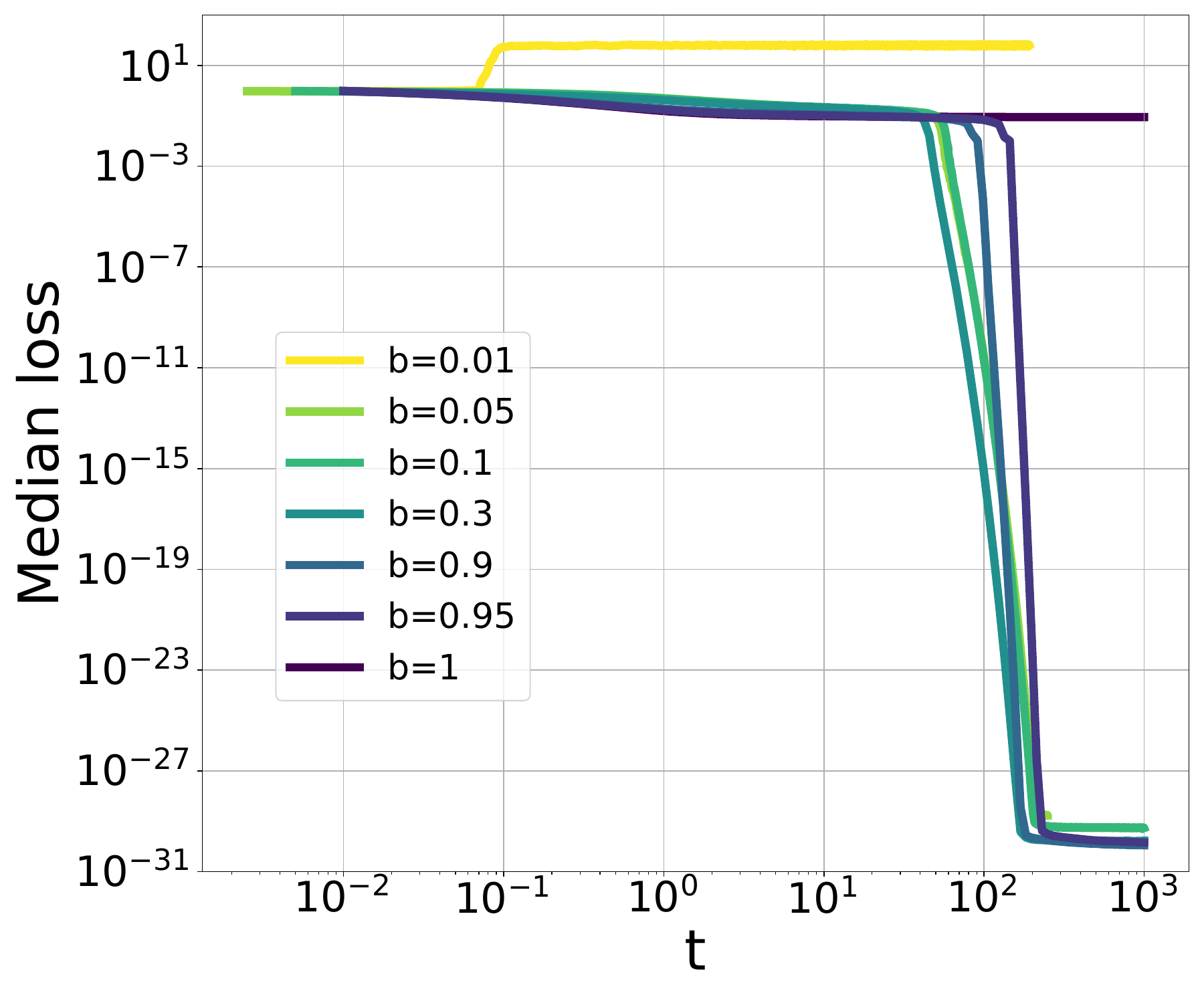}
\hspace{-3mm}
\includegraphics[scale=0.172]{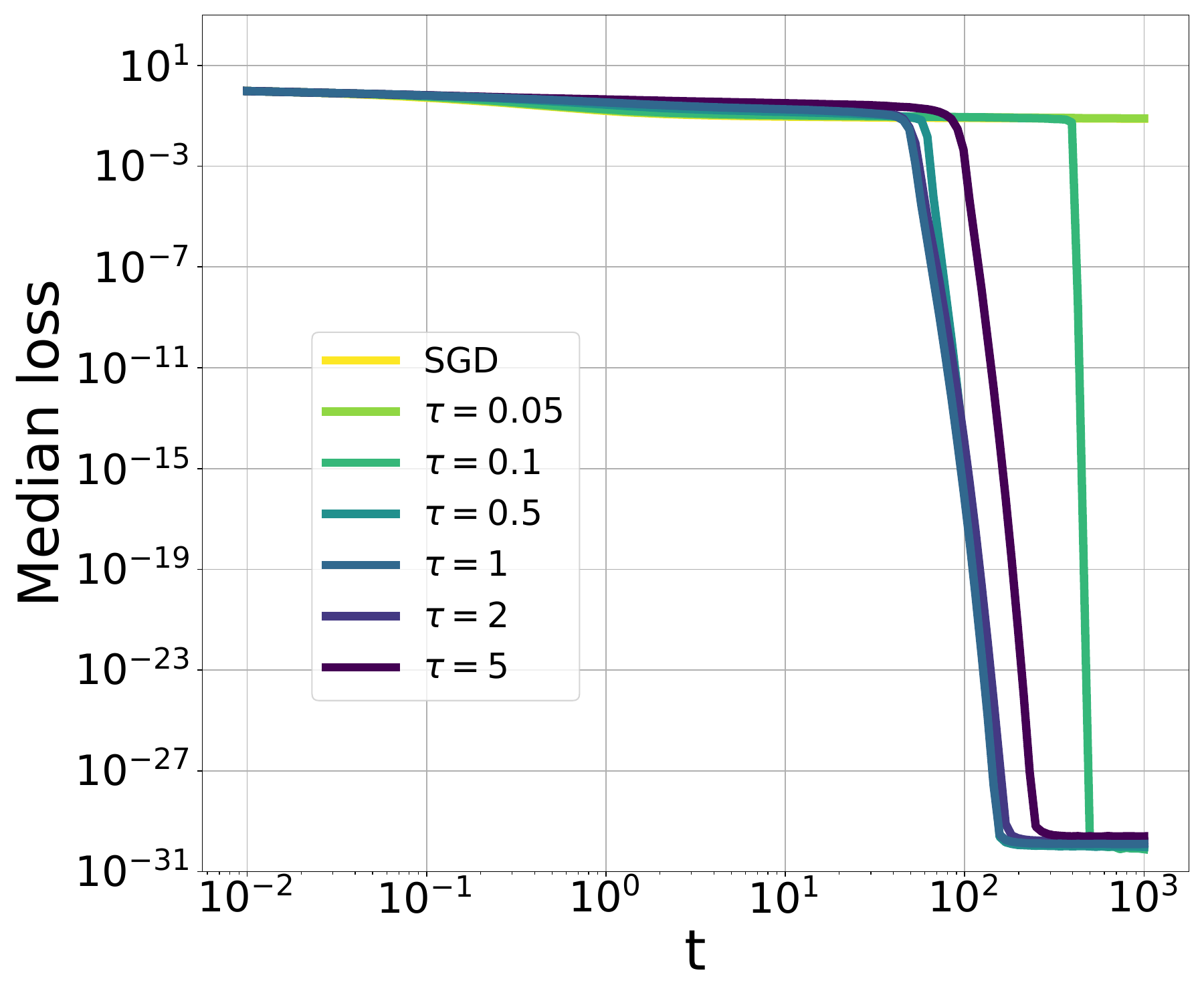}
\hspace{-3mm}
\includegraphics[scale=0.172]{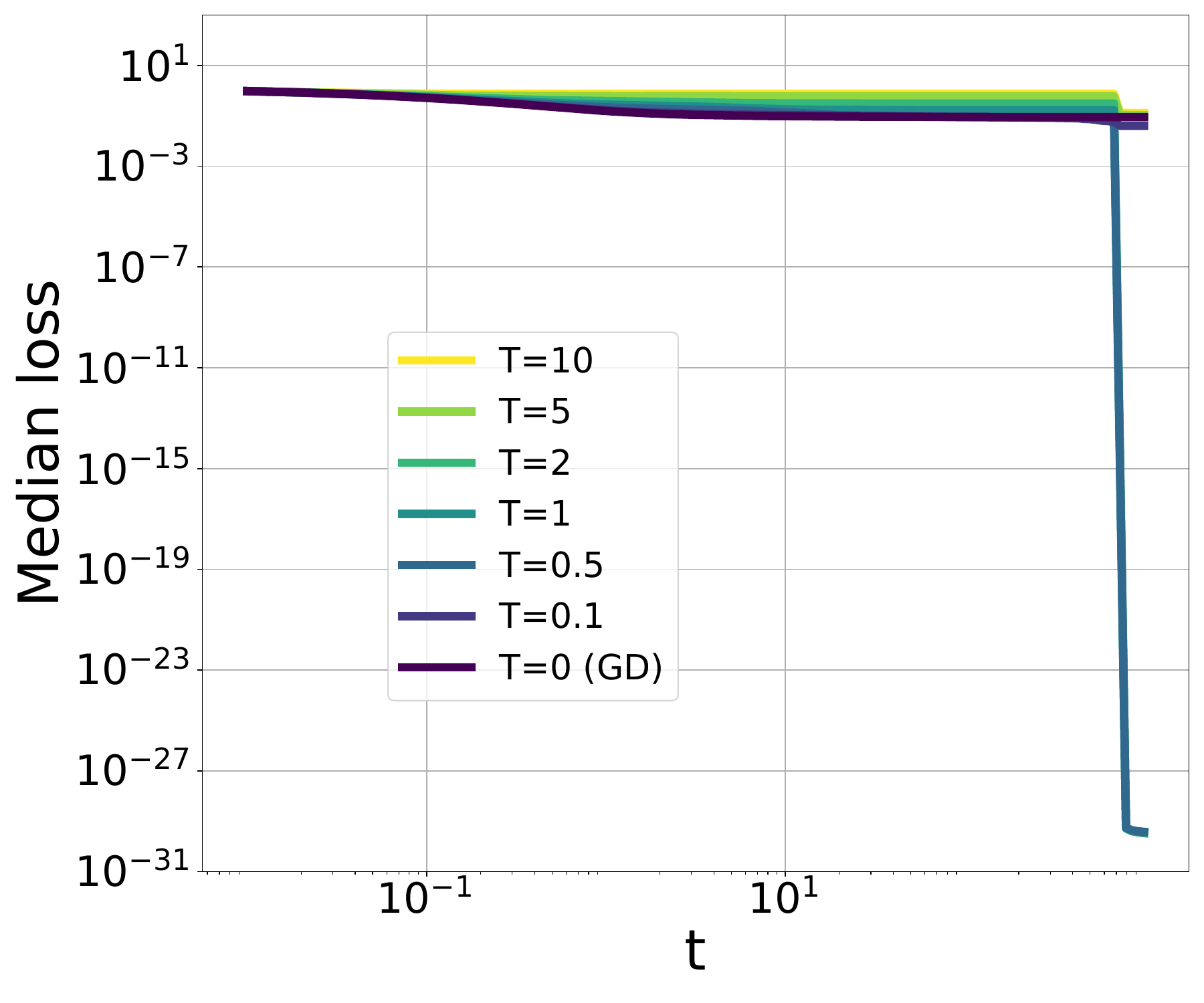}
\end{center}
\caption{\label{tau_vs_temperature} Median magnetization (above) and median loss (below) as a function of training time,  at fixed $\alpha=M/N=3$,  initialization $m_0=0.2$,  dimension $N=1000$, learning rate $\eta=0.01$.  The median is computed over $250$ seeds, drawing a new dataset,  signal and initialization for each seed.  For visibility purposes,  we plot $t+\eta$ on the x-axes.  {\bf Left.} We show persistent-SGD for increasing values of batch size $\b=0.01,0.05,0.1,0.3,0.9,0.95,1$ and fixed persistence time $\tau=1$.  In the case $b=0.1$,  the learning rate has been reduced to $\eta=0.005$,  for $\b=0.01,0.05$ we have used $\eta=0.0025$.  {\bf Center.} We show persistent-SGD at increasing values of persistence time $\tau=0.05,0.1,0.5,1,2,5$ and vanilla SGD, both algorithms at fixed batch size $\b=0.5$.  {\bf Right.} We show the Langevin algorithm at increasing temperatures $T=0$ (GD), $0.1,0.5,1,2,5,10$.  At time $t=700$ the temperature is set to zero. }
\end{figure*}
\paragraph{The role of the noise ---}
Fig.~\ref{tau_vs_temperature} illustrates the effect of different sources of stochasticity on the generalization performance. In particular, we compare the role played by the white noise at temperature $T$ in the Langevin algorithm to the double source of noise in the SGD algorithm: the finite batch size $\b$ and the persistence time $\tau$. 
In the left panel, we depict the dependence of the SGD algorithm on
the batch size, at fixed persistence time. We find that the
generalization performance is non-monotonic in the batch size and the
optimal value is attained at intermediate $\b$. Therefore, at variance
with what observed in deep neural networks trained on real datasets \cite{jastrzkebski2017three,keskar2017largebatch}, in our case we obtain that the optimal batch size is an extensive fraction of the total number of samples.
The central panel displays the (median) performance of SGD for
different values of the persistence time $\tau$, at fixed batch size. 
For times $t\leq \tau$, the samples used to compute the gradient (on
average) do not change, and thus the dynamics presents
plateaus. However, as soon as $t>\tau$, the mini-batch is
refreshed. This results in a sudden increase in performance at times
$t\sim\tau$, that becomes more visible the larger $\tau$. Moreover, we observe a non-monotonic behavior of the
performance as a function of $\tau$. On the one hand, increasing
$\tau$ shifts the final plateau at larger times, delaying the recovery
of the signal. On the other hand, if the persistence time is too
small, the dynamics gets trapped close to the signal, displaying
plateaus followed by sudden jumps similarly as for GD (see
Fig. \ref{SGDvsGD}). There is therefore an intermediate range of
persistence times $\tau$ for which the performance is the best (better
than vanilla SGD). 
Since the literature often compares the SGD noise to the Langevin
noise
\cite{cheng2020stochastic,li2017stochastic,jastrzkebski2017three,hu2017diffusion,anisotropic_noise}
we compare here to the performance achieved by the Langevin algorithm
at fixed temperature. 
The right panel of Fig.~\ref{tau_vs_temperature} depicts the
performance of the Langevin algorithm for different values of
temperature $T$. At large times ($t=700$ in the figure) the
temperature is switched to zero. We find that the best performance is
again reached for intermediate values of the temperature $T$.  We underline the qualitative difference between the effective noise introduced by multi-pass SGD and the white noise of Langevin algorithm. The variance of the noise in Langevin is fixed by the temperature, therefore -- in order to reach a minimum -- an annealing protocol must be implemented and optimized. In contrast, the noise introduced by SGD is automatically reduced during training and it's zero at the global minimum. Therefore, multi-pass SGD has a built-in self annealing protocol, that can be optimized by tuning only two parameters ($\b$ and $\tau$) instead of the whole trajectory of the temperature over time. 
\begin{figure}[!ht]
\centering
\includegraphics[scale=0.266]{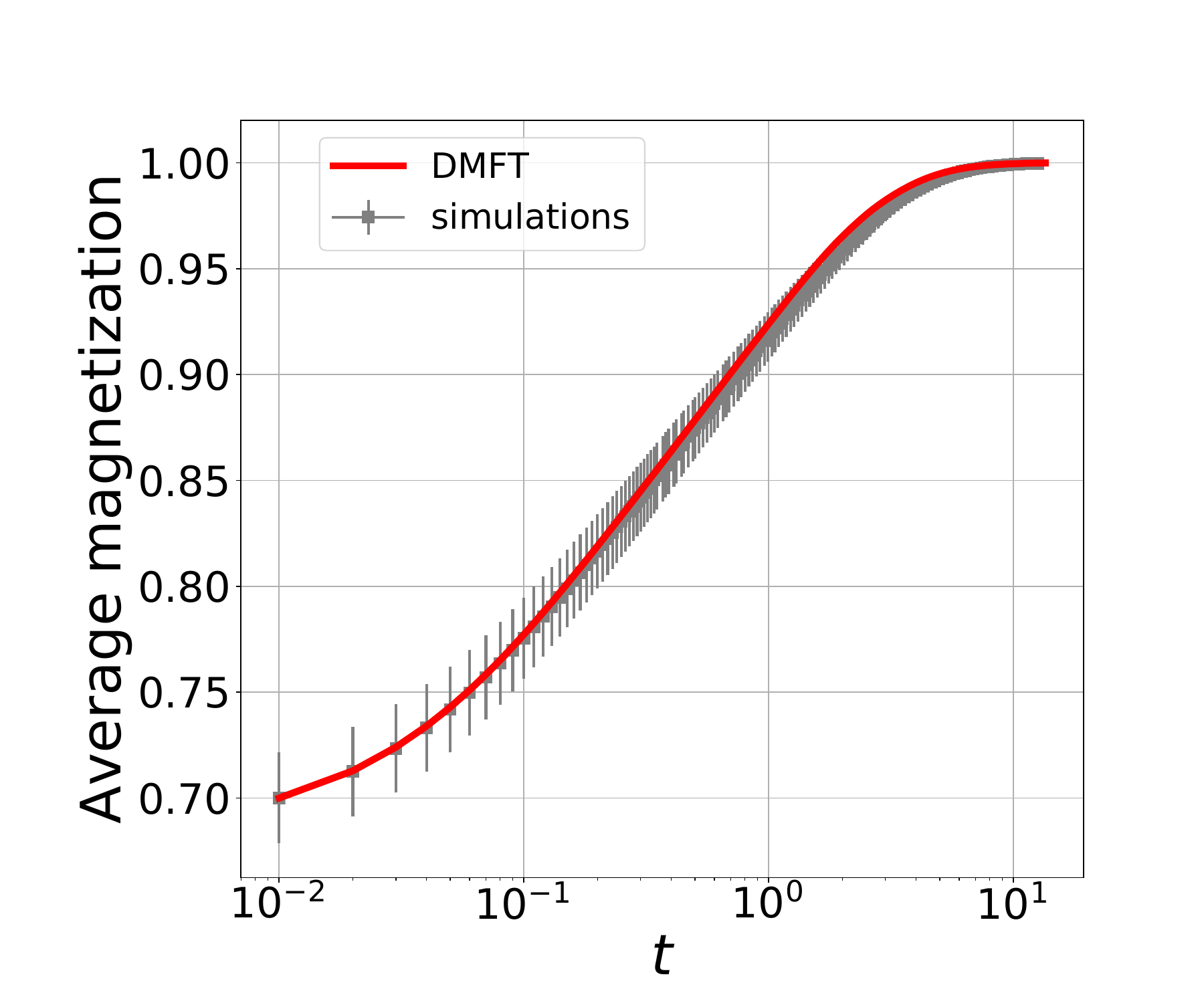}
\hspace{-10mm}
\includegraphics[scale=0.266]{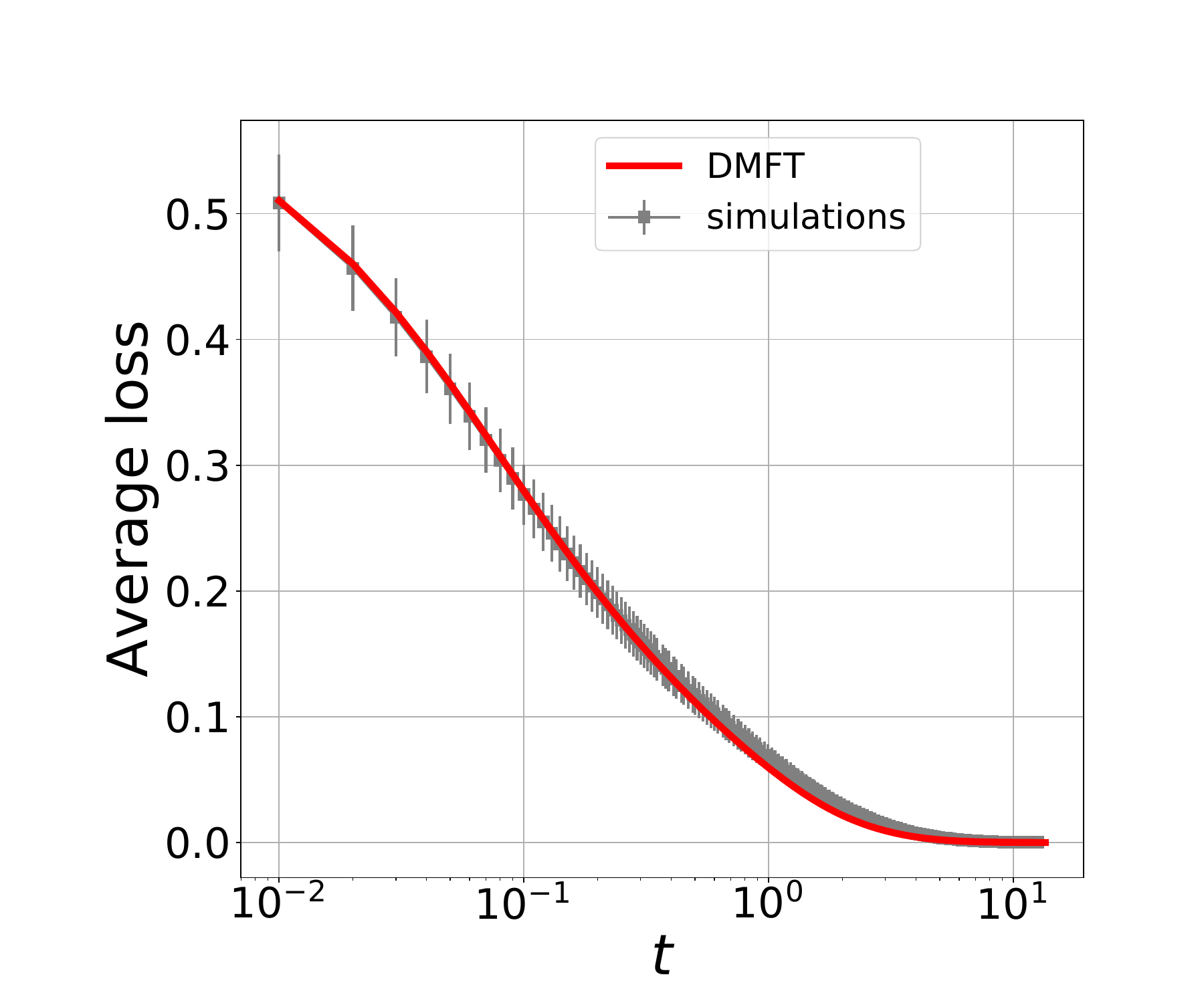}
\caption{\label{DMFT_persistent} Average magnetization (left) and average training loss (right) as a function of time for the persistent SGD algorithm in the spherical setting, at fixed $\alpha=M/N=3$, warm start $m_0=0.7$, persistence time $\tau=2$., batch size $\b=0.6$. The grey dots represent the result from numerical simulations, averaged over $500$ seeds at learning rate $\eta=0.01$ and dimension $N=1000$. The red curve marks the performance predicted by the numerical integration of DMFT equations.}
\end{figure}
\begin{figure}[!ht]
\centering
\includegraphics[scale=0.266]{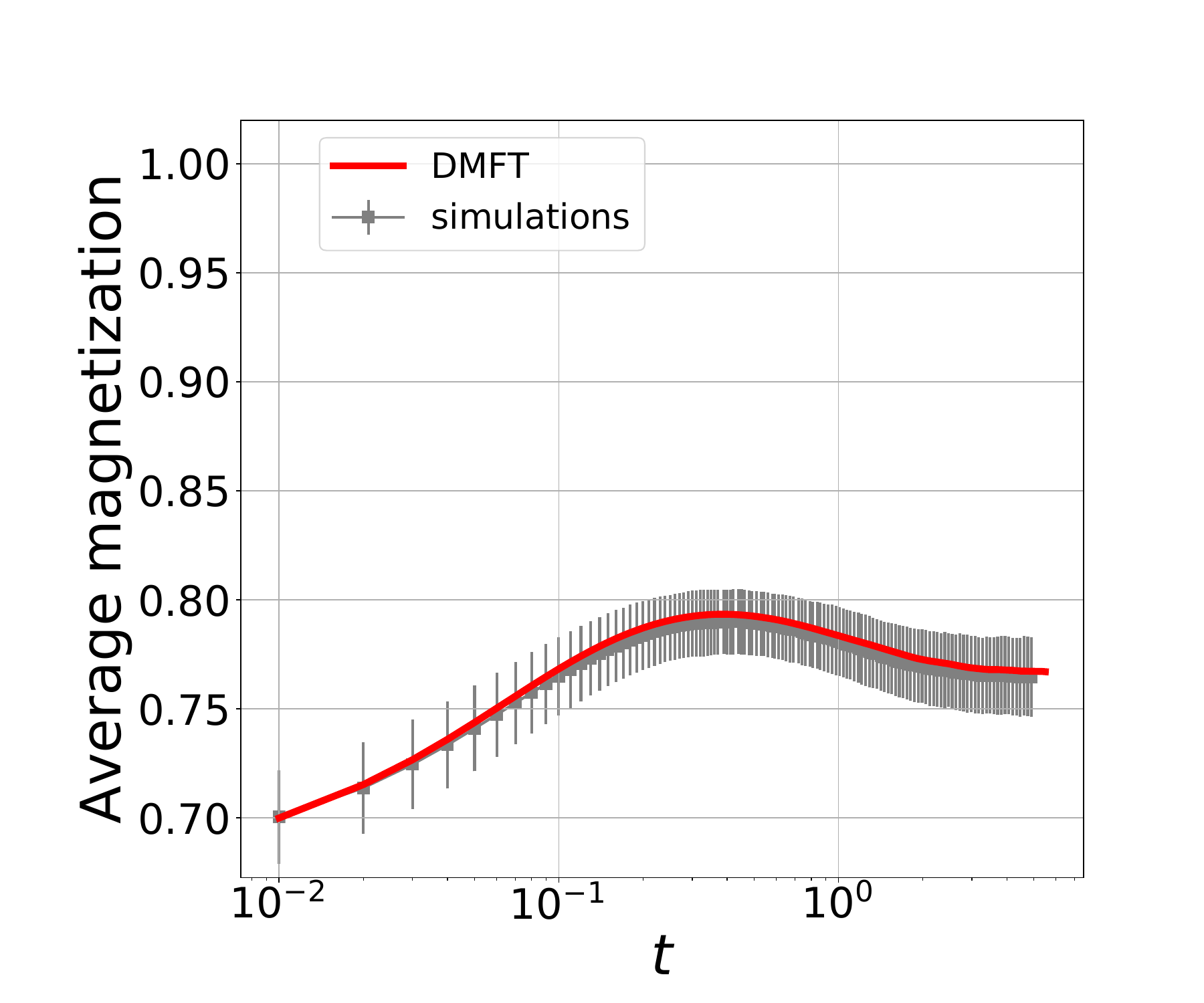}
\hspace{-10mm}
\includegraphics[scale=0.266]{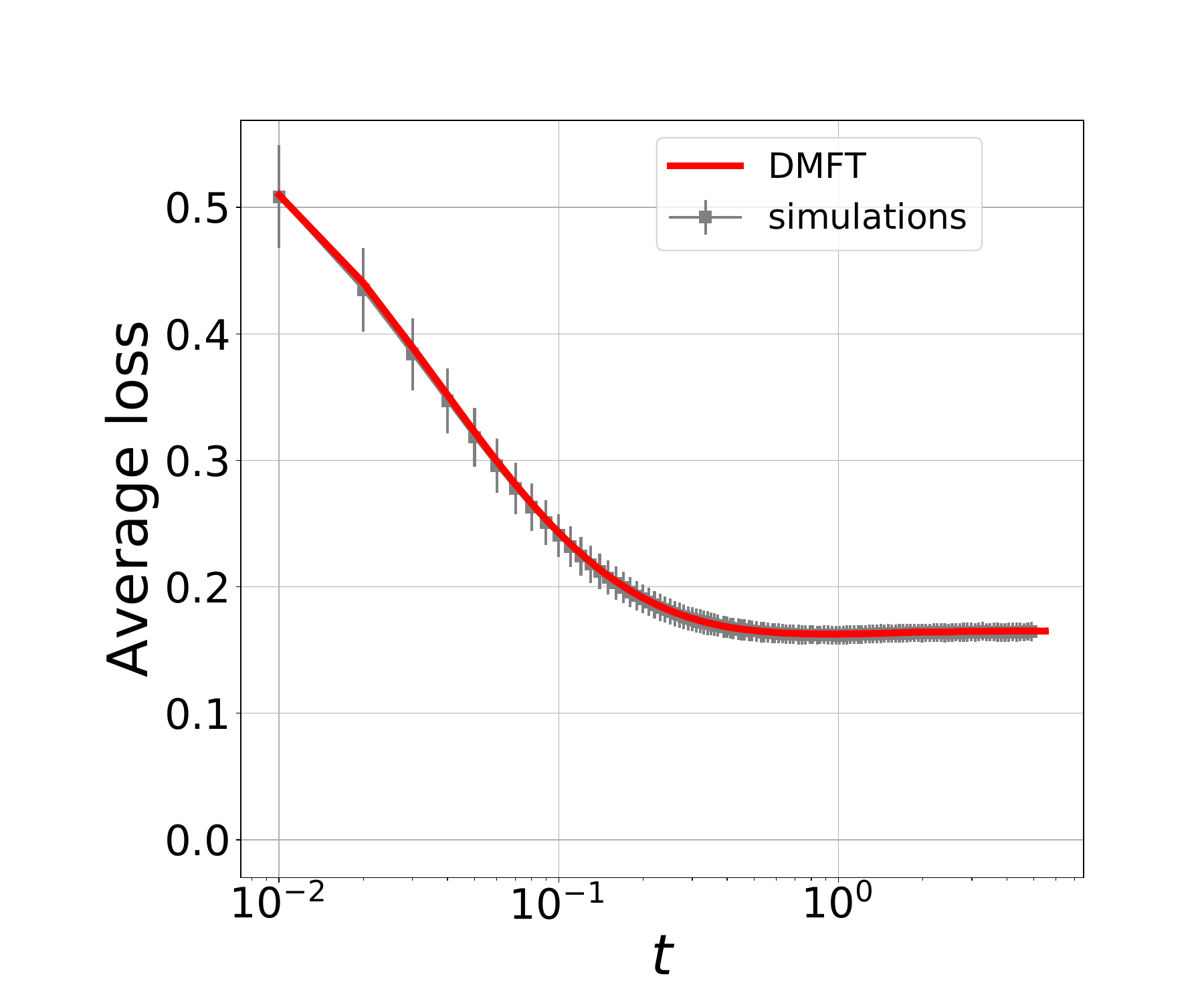}
\caption{\label{DMFT_langevin} Average magnetization (left) and average training loss (right) as a function of time for the Langevin algorithm in the spherical setting, at fixed $\alpha=M/N=3$, warm start $m_0=0.7$, temperature $T=1$. The grey dots represent the result from numerical simulations, averaged over $1000$ seeds at learning rate $\eta=0.01$ and dimension $N=1000$. The red curve marks the performance predicted by the numerical integration of DMFT equations.}
\end{figure}
\paragraph{The analytic characterization}
Fig. \ref{DMFT_persistent} shows the comparison between the average
performance of persistent-SGD obtained from numerical simulations
(grey symbols) with the prediction derived by integrating the DMFT
equations (red line). The left panel depicts the average
magnetization, while the right panel displays the average training loss as a
function of time. Fig. \ref{DMFT_langevin} displays the same
comparison for the Langevin algorithm. In both cases, we find a very
good agreement between theory and simulations.
\begin{figure}[!t]
\centering
\includegraphics[scale=0.266]{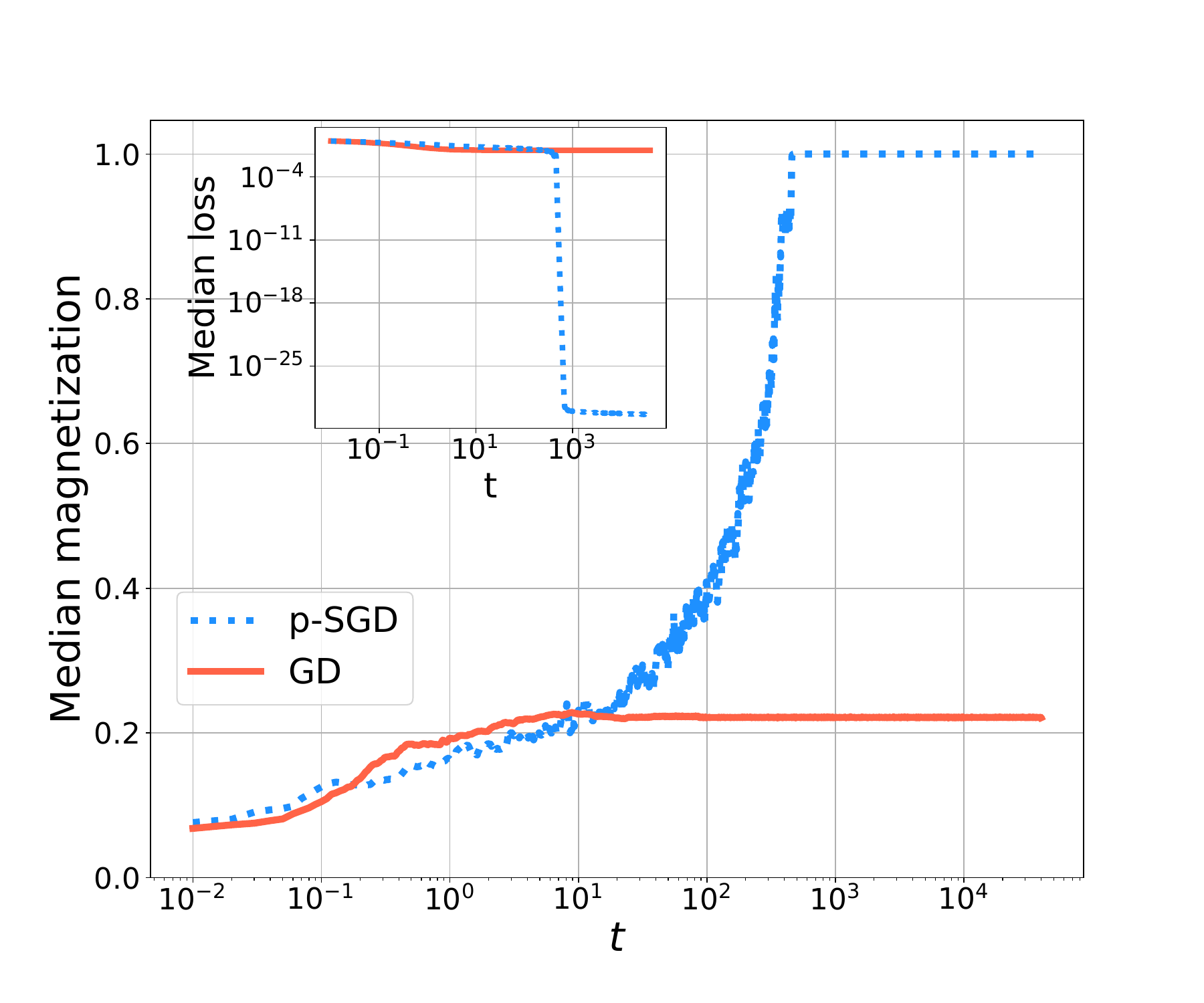}
\hspace{-10mm}
\includegraphics[scale=0.266]{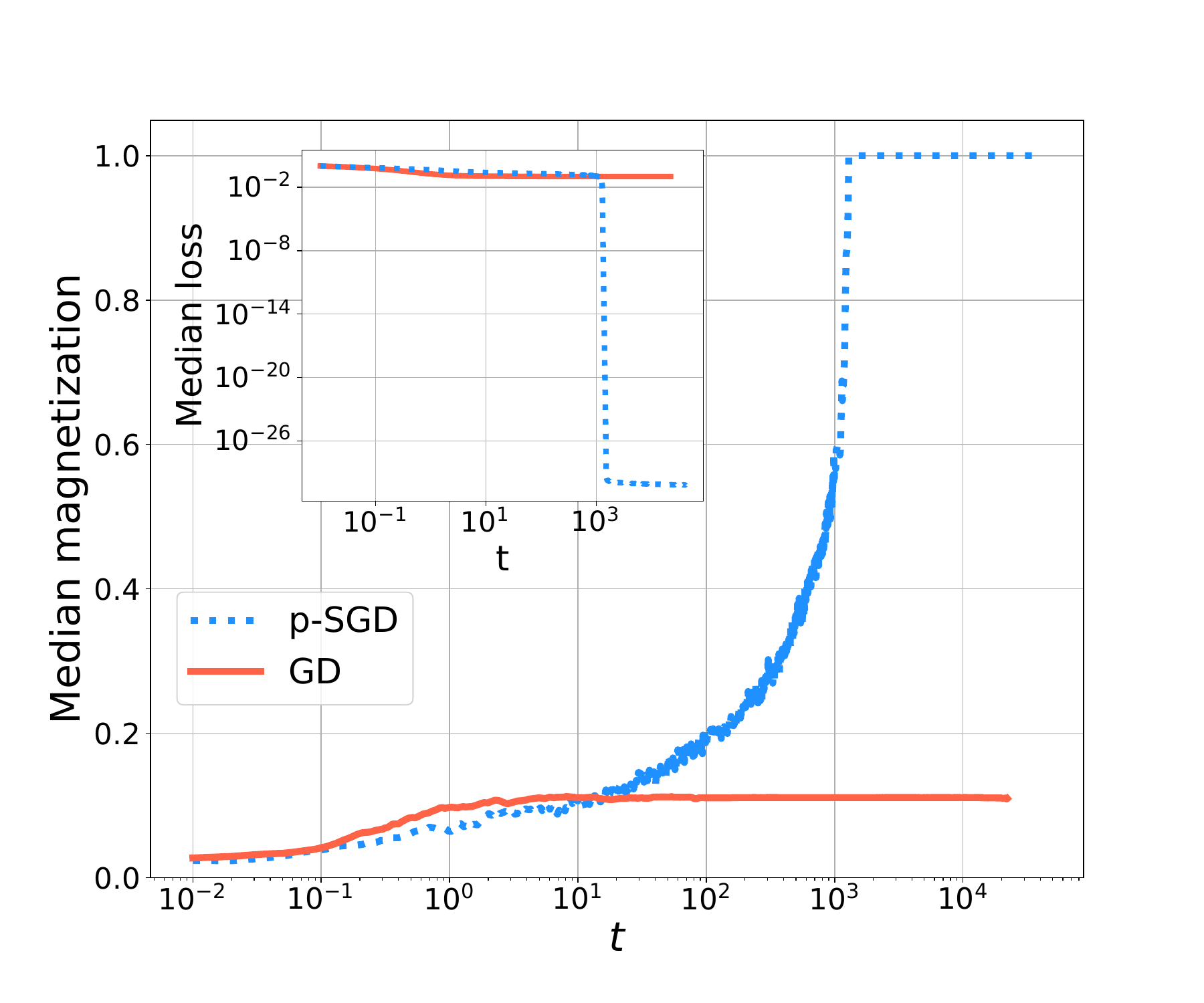}\\
\includegraphics[scale=0.266]{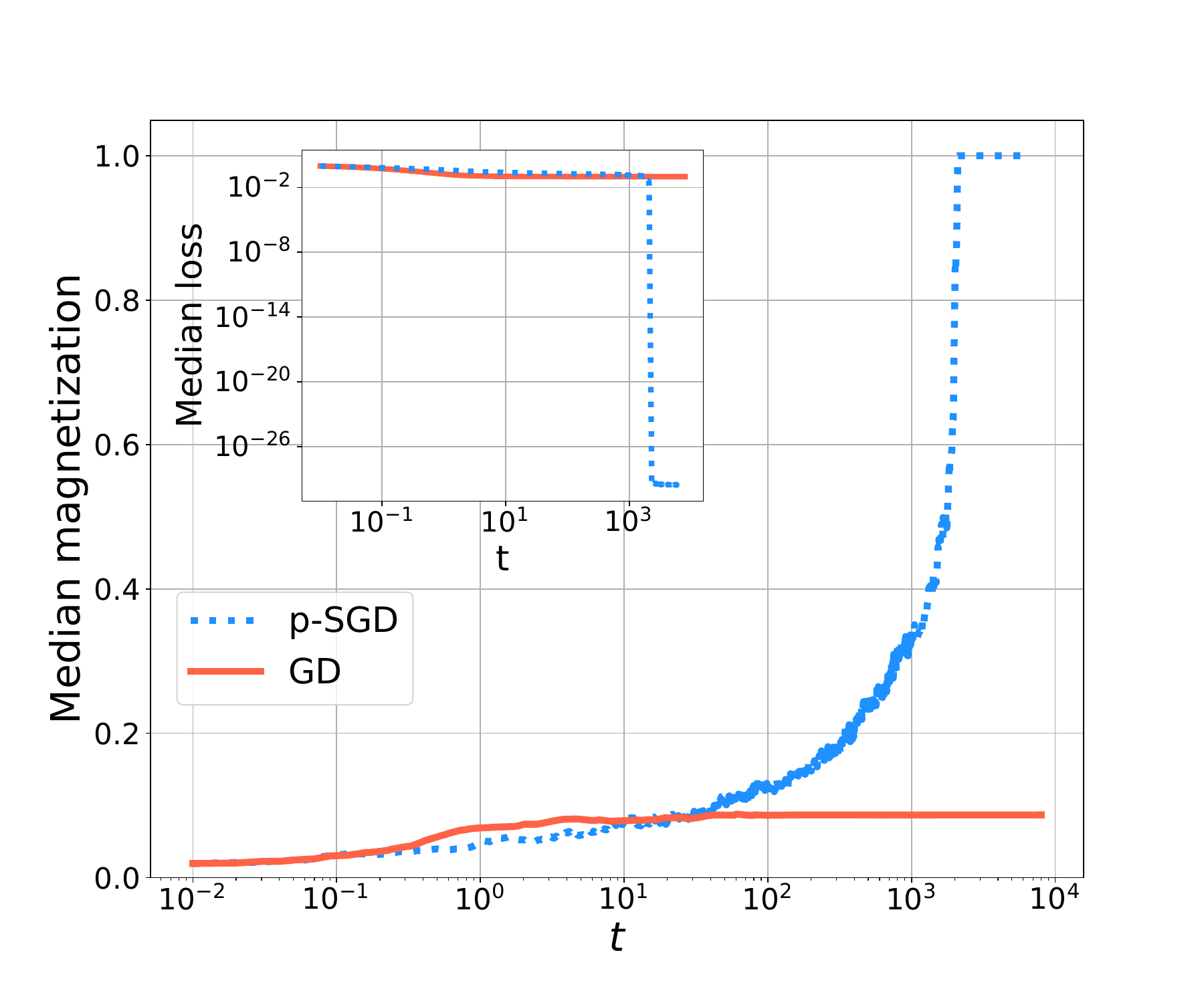}
\hspace{-10mm}
\includegraphics[scale=0.266]{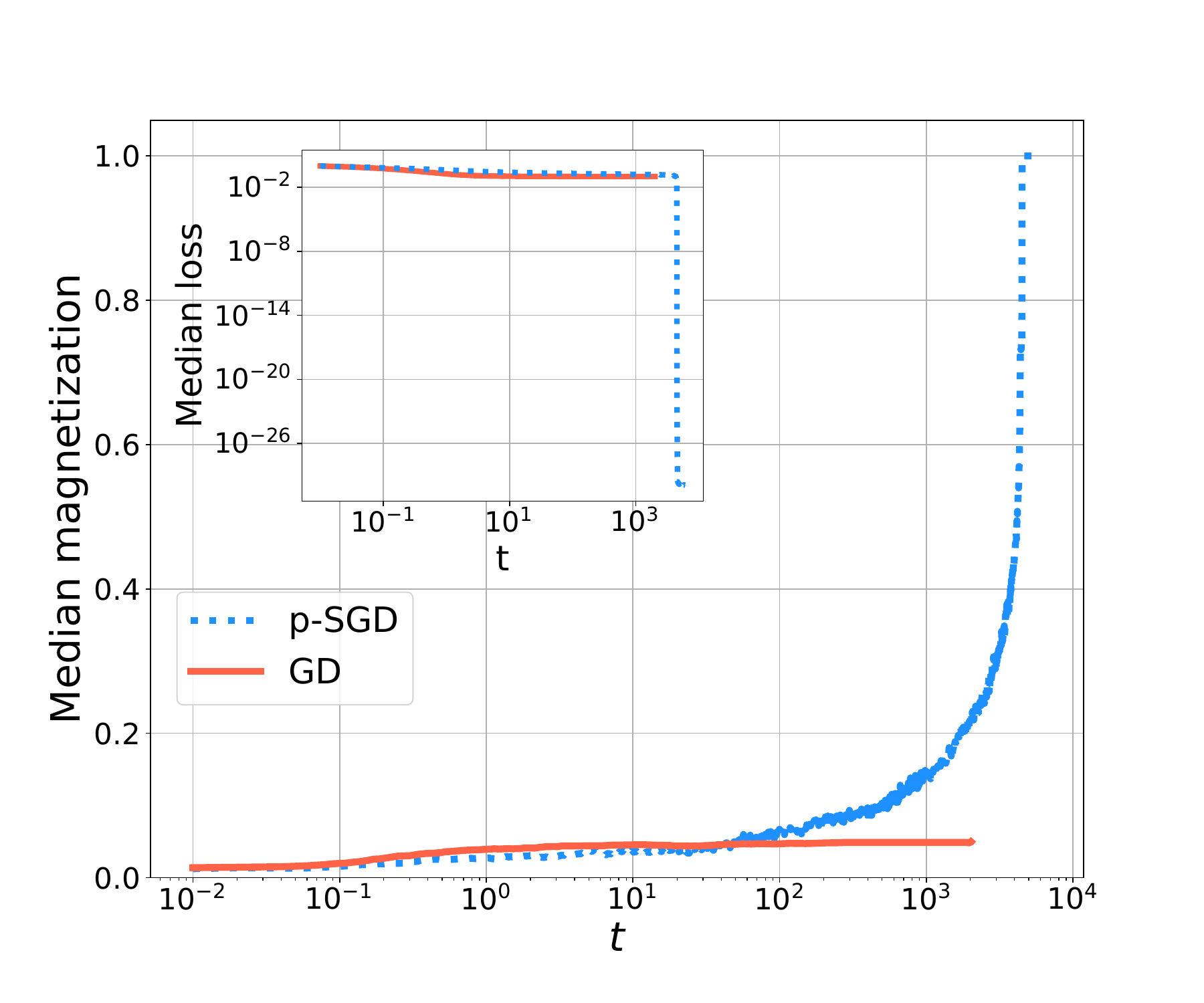}
\caption{\label{fig:random_init} Median magnetization (main plot) and median loss (inset) as a function of time from numerical simulation for the spherical setting at fixed $\alpha=M/N=2.5$, learning rate $\eta=0.01$,  $100$ seeds, and increasing dimension $N=100$ (above-left),  $N=500$ (above-right),  $N=1000$ (below-left),  $N=2500$ (below-right).  We consider random initialization $m_0=0$,  so the finite initial overlap with the signal is only due to finite size effects.  The full red line marks the performance of full-batch gradient descent,  while the dotted blue line represents the persistent-SGD algorithm at batch size $\b=0.5$ and persistence time $\tau=2$.}
\end{figure}
\paragraph{Random initialization}
Fig. \ref{fig:random_init} investigates the behavior of full-batch
gradient descent (full red lines) and persistent SGD (dashed blue
lines) starting from random initialization at fixed
$\alpha=2.5$. Persistent SGD is run at fixed $\b=0.5$, $\tau=2$. We
show the median magnetization (main plots) and the median loss
(insets) as a function of time for increasing values of the dimension:
$N=100$ (above-left panel), $N=500$ (above-right panel), $N=1000$
(below-left panel). and $N=2500$ (below-right panel). In this case
$m_0=0$ and the warm start in the four panels is only given by finite
size effects. We clearly see that, at time scales shown here, gradient
descent is stuck at a plateau of height decreasing as the dimension
$N$ increases. As studied in \cite{mannelli2020}, the recovery
transition of gradient descent starting from random initialization for
comparable system sizes happens at $\alpha \approx 6$, which is few times
larger than the value $\alpha=2.5$  considered here. However, we observe that
already at $\alpha=2.5$ the persistent-SGD algorithm can reach perfect
recovery for the system sizes under consideration. The time to reach
the solution from random initialization is, as expected, compatible with logarithmic
increase in the system size. These observations suggest that the
recovery transition for stochastic gradient descent starting from
random initialization is shifted to lower values of $\alpha$ when
compared to gradient descent. This is an interesting direction for future investigations.
\section{Conclusion}
In this paper, we have considered the real-valued phase retrieval problem as a paradigmatic highly non-convex optimization problem to test the generalization performance of full-batch  
gradient descent and some of its stochastic variants: multi-pass SGD,
its persistent version, and the Langevin algorithm. We have shown that
stochasticity is crucial to achieve perfect recovery of the hidden
signal at low sample complexity so that stochastic gradient descent
outperforms gradient descent in this task. We have observed intriguing
features of the loss profile and illustrated how various sources of
noise allow the dynamics to circumvent the traps in the landscape. We
have provided an analytic description of the learning curve in the
infinite-dimensional and continuous-time limit via the dynamical
mean-field theory, showing that the observed behavior is not due to finite size effects or to a finite learning rate. The present work leads to interesting extensions both on the analytic and numerical sides. On the one hand, the characterization of the dynamical evolution of the algorithms via DMFT can be extended to include realistic initializations (e.g., spectral initialization). On the other hand, it would be interesting to test the persistent variant of multi-pass SGD and investigate the role of the persistence time on real datasets and architectures, which we leave for future work.
\section{Acknowledgements}
We thank Stefano Sarao Mannelli for useful discussions.  This work was
supported by ERC under the European Union’s Horizon
2020 Research and Innovation Programme Grant Agreement 714608-SMiLe,
and by ''Investissements d'Avenir'' LabExPALM (ANR-10-LABX-0039-PALM).
\clearpage
\appendix

\section{Derivation of DMFT equations}
\label{sec:derivationDMFT} 
In this section, we provide additional details on the derivation and numerical integration of the theoretical equations describing the learning performance via dynamical mean-field theory (DMFT), presented in Sec. \ref{sec:analytic_characterization} of the main text. 
The computation is on the line of the one presented in \cite{ABUZ18,francesca2020dynamical}. Here we consider a different loss function, i.i.d. Gaussian input data and labels generated by a teacher vector. We have to take into account the spherical constraint and the additional white noise of the Langevin algorithm. We use the Martin-Siggia-Rose-Janssen-deDominicis (MSRJD) path-integral formalism. 
We start by writing the dynamical partition function 
\beq
\begin{split}
1&=Z_{\rm dyn}=\int \DD \underline w \prod_{i=1}^N \delta\left(-\frac{\partial w_i(t)}{\partial t} -\hat \nu(t)w_i(t) - \frac{\partial }{\partial w_i}\frac 1\b\sum_{\mu=1}^{\a N} s_{\mu}(t)v(h_\mu(t);h_\mu^{(0)} ) +\varsigma_i(t)\right)\\
&=\int \DD \underline{w} \,\DD\hat{\underline w}\\&\times\exp\left[\int\dde t\,  i\hat{\underline w}(t)\cdot \left(-\frac{\partial \underline w(t)}{\partial t} -\hat \nu(t)\underline w(t) - \frac{\partial }{\partial \underline w}\frac 1\b \sum_{\mu=1}^{\a N} s_{\mu}(t)v(h_\mu(t);h_\mu^{(0)} ) + \underline{\varsigma}(t)\right)\right].
\end{split}
\eeq
Since the dynamical partition function is strictly equal to one, we can safely average its expression on the random patterns $\underline \xi^\mu$ and on the Langevin noise $\underline{\varsigma}$.
Following \cite{ABUZ18} we can use a supersymmetric formalism to proceed in a compact way.
In this case the dynamical variables $\underline w(t)$ are merged with their auxiliary fields $ \underline{\hat w}(t)$ in a superfield involving a couple of Grassmann variables $\theta_a, \bar\theta_a$ \cite{Zi96}:
\beq
\underline w(a) = \underline w(t) +i \th_a \overline \th_a  \underline{\hat w}(t).
\eeq
In this way the dynamical partition function (averaged over the Langevni noise) can be written as
\beq
Z_{\rm dyn}=\int \DD \underline w(a) \exp\left[-\frac 12 \int \dde a \dde b \,\underline w(a) \KK (a,b) \underline w(b) +\alpha N\ln \ZZ\right],
\eeq
where
\beq
\ZZ= \int \DD h(a) \DD \hat h(a) \int \dde h_0 \dde \hat h_0\exp\left[ \SS_{\rm loc}\right],
\eeq
and the kinetic kernel $\KK(a,b)$ is implicitly defined in such a way that
\beq
-\frac 12 \int \dde a\dde b \,\underline w(a)\KK(a,b)\underline w(b) = -i\int \dde t\, \hat{\underline w}(t)\cdot\left( \frac{\partial \underline w(t)}{\partial t}+\hat \nu(t)\underline w(t)-iT\underline{\hat w}(t)\right).
\eeq
In particular, we have
\begin{equation}
\begin{split}
\mathcal{K}(a,b)&=-2T\delta(t_a-t_b)-\theta_a\bar\theta_a \partial_{t_a}\delta(t_b-t_a)-\theta_b\bar\theta_b \partial_{t_b}\delta(t_a-t_b)+\hat{\nu}(a)\delta(a,b),\\
\hat\nu(a)&=\hat\nu(t_a),\\
\delta(a,b)&=\delta(t_a-t_b)(\theta_a\bar\theta_a-\theta_b\bar\theta_b).
\end{split}\label{Kab}
\end{equation}
The local action $\SS_{\rm loc}$ is defined as
\beq
\begin{split}
\SS_{\rm loc} &= ih_0\hat h_0 + i \int \dde a \,\hat h(a) h(a) -\frac 12 \left[\hat h_0^2 +2 \hat h_0 \int \dde a \,\hat h(a) m(a) +\int \dde a \dde b\, \hat h(a) \hat h(b) Q(a,b)\right] \\
&- \frac 1\b\int \dde a\, s(a)v(h(a); h_0),
\end{split}
\eeq
where $s(a)=s(t_a)$, and we have introduced the dynamical order parameters 
\beq
m(a) = \frac 1 N  \underline w(a) \cdot \uwo ,\ \ \ \ \ \ \ Q(a,b) = \frac 1N \underline w(a)\cdot \underline w(b).
\eeq
Performing the Gaussian integration over the superfields $\underline w(a)$, we obtain
\beq
\overline {Z_{\rm dyn}} = \int \DD Q(a,b)\DD m(a)\, e^{N\AA_{\rm dyn}}
\eeq
where
\beq
\AA_{\rm dyn} = -\frac{1}{2}\int \dde a \dde b\, \KK(a,b)\left[Q(a,b)+m(a)m(b)\right]+\frac 12 \ln \det Q + \alpha \ln \ZZ_{\rm loc}\label{Adyn}
\eeq
where
\beq
\begin{split}
\ZZ_{\rm loc} = \int \frac{\dde h_0}{\sqrt{2\pi}}e^{-h_0^2/2} \int \DD h(a) \, \exp\left[-\frac 12 \int \dde a \dde b \,h(a) Q^{-1}(a,b) h(b) \right.\\\left.-\frac 1\b\int \dde a\, s(a)v(h(a)+h_0 m(a);h_0)\right].
\end{split}
\eeq
At this point we can evaluate the integral over $Q$ and $m$ through a saddle point computation.
The saddle point equations read
\begin{eqnarray}
0=&-\KK(a,b) + Q^{-1}(a,b) +2\a \frac{\delta \ln \ZZ_{\rm loc}}{\delta Q(a,b)}\label{speQ}\\
0=&-\int \dde b\,\KK(a,b)m(b) +\a \frac{\delta \ln \ZZ_{\rm loc}}{\delta m(a)}\label{speM}
\end{eqnarray}
We can evaluate the functional derivatives
\beq
\begin{split}
2\alpha \frac{\delta \ln \ZZ_{\rm loc}}{\delta Q(a,b)} &= \frac{\a}{\b^2} \langle s(a)s(b)\partial_1 v(h(a)+h_0 m(a);h_0)\partial_1v(h(b)+h_0 m(b);h_0)\rangle \\&\quad- \frac \a \b\delta(a,b) \langle s(a)\partial_1^2 v(h(a)+h_0m(a);h_0)\rangle
\equiv M(a,b) - \delta \nu(a)\d(a,b),\\
\alpha \frac{\delta \ln \ZZ_{\rm loc}}{\delta m(a)} &= -\frac \a \b\langle h_0s(a)\partial_1 v(h(a)+h_0 m(a);h_0)\rangle,
\end{split}
\eeq
where we have defined
\beq
\begin{split}
M(a,b) & =\frac{\a}{\b^2} \langle s(a)s(b)\partial_1 v(h(a)+h_0 m(a);h_0)\partial_1v(h(b)+h_0 m(b);h_0)\rangle,\\
\delta \nu(a) &=\frac \a \b \langle s(a) \partial_1^2 v(h(a)+h_0m(a);h_0)\rangle.
\end{split}
\eeq
The average in brackets denotes the average with the following measure
\beq
\begin{split}
\langle \bullet \rangle = \int \frac{\dde h_0}{\sqrt{2\pi}} \int \DD h(a) \, \bullet \\\times\exp\left[-\frac{h_0^2}{2}-\frac 12 \int \dde a \dde b \,h(a) Q^{-1}(a,b) h(b) -\frac 1\b \int \dde a\, s(a)v(h(a)+h_0 m(a);h_0)\right].\label{Seff}
\end{split}
\eeq
Along the lines of \cite{ABUZ18}, we can rewrite the average in Eq. \eqref{Seff} as an average over $h_0\sim \mathcal{N}(0,1)$, the variables $s(t)$ defined in Eq. \eqref{persistentSGD} of the main text, and an effective stochastic process
\beq
\partial_t h(t) = - \tilde \nu(t) h(t) - \frac 1\b s(t)\partial_1v(\tilde h(t);h_0) + \int_0^t \dde t'\, M_R(t,t') h(t') + \chi(t),\label{supmat_eff_process}
\eeq
with initial condition $P(h(0)) = e^{-h(0)^2/2(1-m_0^2)}/\sqrt{2\pi (1-m_0^2)}$, where $\chi(t)$ is an effective Gaussian noise:
\beq 
\langle\chi(t) \rangle = 0 ,\ \ \ \ \ \ \ \ \langle \chi(t) \chi(t')\rangle =2T\delta(t-t') +M_C(t,t')
\eeq
and we have defined the following auxiliary functions:
\begin{equation}
\begin{split}
\tilde h(t) &\equiv h(t) + h_0 m(t),\\
\delta\nu(t)&=\frac \alpha \b \langle s(t) \partial_1^2v(\tilde h(t);h_0)\rangle ,\\
\hat \nu (t)&=-\frac \alpha \b \langle s(t)\tilde h(t) \partial_1 v(\tilde h(t);h_0)\rangle +T,\\
\tilde \nu(t) &= \hat\nu(t) + \delta\nu(t).\label{Nu}
\end{split}
\end{equation}
The expression for the Langrange multiplier $\hat \nu(t)$ is obtained enforcing the spherical constraint $\sum_{i=1}^N {\rm d}w_i^2/{\rm d}t=0$ by applying It\^{o}'s formula to Eq. \eqref{continuous_equation} of the main text. The kernels $M_C(t,t')$ and $M_R(t,t')$ are obtained expanding $M(a,b)$:
\begin{equation}
\begin{split}
M(a,b)&=M_C(t_a,t_b)+\theta_a\bar\theta_a M_R(t_b,t_a)+\theta_b\bar\theta_b M_R(t_a,t_b),\\
M_C(t,t') &= \frac{\a}{\b^2}  \langle s(t)s(t') \partial_1 v(\tilde h(t);h_0)\partial_1 v(\tilde h(t');h_0)) \rangle,\\
M_R(t,t')&=\frac{\alpha}{\b^2} \langle s(t)s(t')  \partial_1v(\tilde h(t);h_0)\partial_1^2 v(\tilde h(t');h_0)\,i\hat h(t')\rangle\\&=\frac{\alpha}{\b^2} \frac{\delta}{\delta P(t')}\langle s(t)\partial_1  v(\tilde h(t);h_0)\rangle\biggr\rvert_{P=0}.
\end{split}\label{supmat_kernels}
\end{equation}
The variable $P(t')$ indicates a linear perturbation applied on the gap variable $h$ at time $t'$ and then set to zero. The kernel $M_R(t,t')$ can be also expressed as 
\beq
M_R(t,t')= \frac{\alpha}{\b^2} \langle s(t) \partial_1^2 v(\tilde h(t);h_0) T(t,t') \rangle,
\eeq
where $T(t,t')=\delta h(t)/\delta P (t')$ satisfies
\beq
\partial_t T(t,t')=-\tilde\nu (t) T(t,t') -  \frac 1\b s(t)\partial_1^2 v(\tilde h(t);h_0) \left(T(t,t')-\delta(t,t')\right)+\int_{t'}^t \dde s\, M_R(t,s) T(s,t').\label{Ttt}
\eeq
Furthermore, from Eq. \eqref{speM} we get the behavior of the magnetization $m(t)$ as a function of time:
\beq 
\partial_tm(t) = -\hat \nu(t) m(t) - \mu(t),\ \ \ \ \ \ \ \ \ \ \ m(0)=m_0,\label{evolution_m}
\eeq
where $m_0$ is defined in Eq. \eqref{init} of the main text and 
\beq
\mu(t)=\frac \alpha \b\langle s(t) h_0 \partial_1 v(\tilde h(t);h_0)\rangle.
\eeq
Moreover, setting $m(t)=0$ one gets a set of equations that coincide with \cite{ABUZ18}. From the solution $Q(a,b)$ of the saddle point Eq. \eqref{speQ}, we can obtain the equations for the dynamical correlation function $C(t,t')=\sum_i w_i(t)w_i(t')/N$ and the response $R(t,t')=\sum_i\delta w_i(t)/\delta H_i(t')/N$ to a linear perturbation of the weights by an infinitesimal local field $H_i(t)$. Indeed, we can write the closure relation
\beq
\begin{split}
\delta(a,b)&=\int \dde c \,Q^{-1}(a,c)Q(c,b)\\
&=\int \dde c\, \left[\mathcal{K}(a,c)-\mathcal{M}(a,c)\right]Q(c,b) + \delta\nu(a)Q(a,b).\label{closure}
\end{split}
\eeq
Now we can express the overlap explicitly in time and Grassman coordinates
\beq
\begin{split}
Q(a,b)=\frac{1}{N}\underline{w}(a)\cdot\underline{w}(b)=C(t_a,t_b)-m(t_a)m(t_b)+\theta_a \bar\theta_a R(t_b,t_a)+\theta_b \bar\theta_b R(t_a,t_b),
\end{split}
\eeq
where we remind that in \eqref{Adyn} we have performed the change of variable $Q(a,b)\rightarrow Q(a,b)+m(a)m(b)$. Plugging Eq. \eqref{Kab} and Eq. \eqref{supmat_kernels} in Eq. \eqref{closure}, we find
\begin{equation}
\begin{split}
\delta(t_a-t_b)(\theta_a\bar\theta_a-\theta_b\bar\theta_b)=\\-2TR(t_b,t_a)+\partial_{t_a}C(t_a,t_b)-\partial_{t_a}m(t_a)m(t_b)+\hat\nu(t_a)\left(C(t_a,t_b)-m(t_a)m(t_b)\right)\\-\int \dde t_c \,\left[ M_C(t_a,t_c)R(t_b,t_c)+M_R(t_a,t_c)\left(C(t_b,t_c)-m(t_b)m(t_c)\right)\right]\\+\theta_a\bar\theta_a\left[\partial_{t_a}R(t_b,t_a)+\hat\nu(t_a)R(t_b,t_a)\right]-\theta_a\bar\theta_a\int \dde t_c \,M_R(t_c,t_a)R(t_b,t_c)\\+\theta_b\bar\theta_b\left[\partial_{t_a}R(t_a,t_b)+\hat\nu(t_a)R(t_a,t_b)-\int \dde t_c\,M_R(t_a,t_c)R(t_c,t_b)\right]
\\+\delta\nu (t_a)\left(C(t_a,t_b)-m(t_a)m(t_b)+\theta_a \bar\theta_a R(t_b,t_a)+\theta_b \bar\theta_b R(t_a,t_b)\right).
\end{split}
\end{equation}
We can derive two equations from the scalar and Grassman terms (the terms in $\theta_a\bar\theta_a$ and $\theta_b\bar\theta_b$ result in the same contribution):
\beq 
\begin{split}
\partial_t C(t',t)=&-\tilde\nu(t)C(t,t')+2TR(t',t)+\int_0^t \dde s \,M_R(t,s) C(t',s)+\int_0^{t'}\dde s \, M_C(t,s)R(t',s)
\\
&-m(t')\left(\int_0^t \dde s \, M_R(t,s)m(s)+\mu(t)-\delta\nu(t)m(t)\right) \ \ \ \ \  {\rm if \,\,} t\neq t'  ,\\
\partial_t R(t,t')=&-\tilde\nu(t)R(t,t')+\delta(t-t')+\int_{t'}^t \dde s\, M_R(t,s)R(s,t'),
\end{split}\label{CR}
\eeq
where we have used Eq. \eqref{evolution_m} in the first of Eqs. \eqref{CR} . 
An alternative expression to Eq. \eqref{Nu} for the Lagrange multiplier $\hat\nu(t)$ can be obtained by plugging $C(t,t)=1$ in the first of eqs. \eqref{CR}:
\beq
\begin{split}
\hat\nu(t)= &-\delta\nu(t) +T+ \int_0^t \dde s \left(  M_R(t,s) C(t,s) + M_C(t,s)R(t,s)\right) \\&-m(t)\left(\mu(t)-\delta\nu(t)m(t) +\int_0^t \dde s M_R(t,s)m(s)\right).\label{hatnu2}
\end{split}
\eeq

\subsection{Numerical integration of the DMFT equations}
\label{sec:numericsDMFT}
In this section, we provide more details on the numerical integration of DMFT equations. Similarly as in \cite{francesca2020dynamical}, we implement an iterative scheme to reach the convergence of the self-consistent process in Eq. \eqref{supmat_eff_process}:
\begin{itemize}
\item We start from a simple guess of the kernels in Eq. \eqref{supmat_kernels} and the auxiliary functions in \eqref{Nu}. In particular, we set $m(t)=m_0\,\, \forall t$, $M_R(t,t')=0 \,\,\forall t,t'$, $M_C(t,t)=M_C(0,0)\,\,\forall t$, $M_C(t,t')=0.1\times M_C(0,0)\,\,\forall t\neq t'$, and we initialize all the entries of $\tilde{\nu}(t)$ and $\mu(t)$ to their value at $t=0$. 
\item We use the previous guess to generate multiple realizations of the curve $h(t)$.
\item We update the kernels and auxiliary functions, computing the averages over $h(t)$, $h_0$, $s(t)$. We introduce a damping in the update to control the oscillations. We integrate Eq. \eqref{evolution_m} to obtain the magnetization $m(t)$.
\item We repeat the above procedure until the kernels and auxiliary functions reach a fixed point. 
\end{itemize}
We use Eq. \eqref{hatnu2} in order to compute the Lagrange multiplier $\hat \nu (t)$ because we find that it is more stable to fluctuations. We integrate Eq. \eqref{Ttt} to compute the kernel $M_R$. We typically use a discrete time step $dt=10^{-3}-10^{-2}$.

\section{Generalization error}
\label{sec:generalization_error}
\begin{figure}
\centering
\includegraphics[scale=0.181]{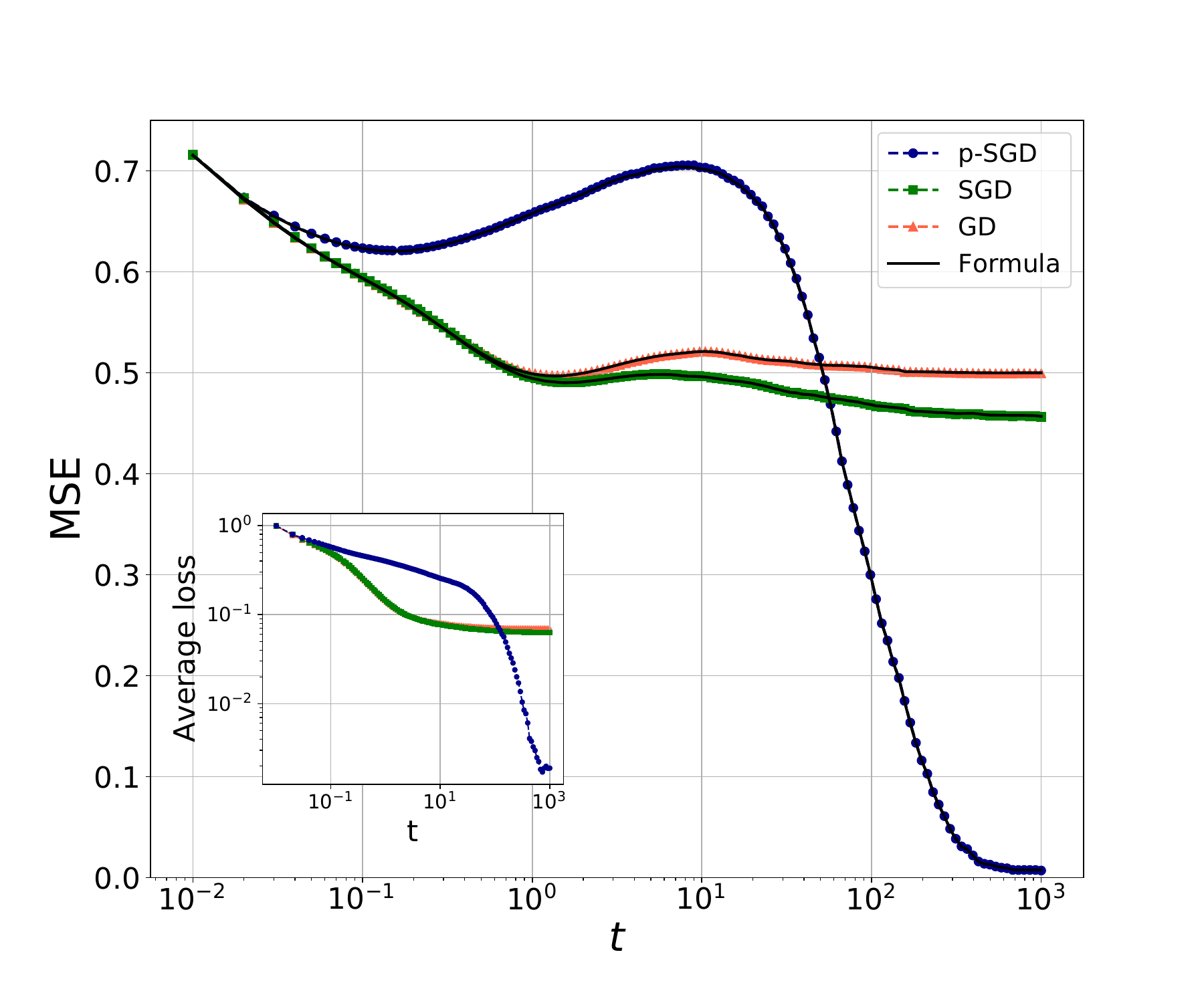}
\hspace{-7mm}
\includegraphics[scale=0.181]{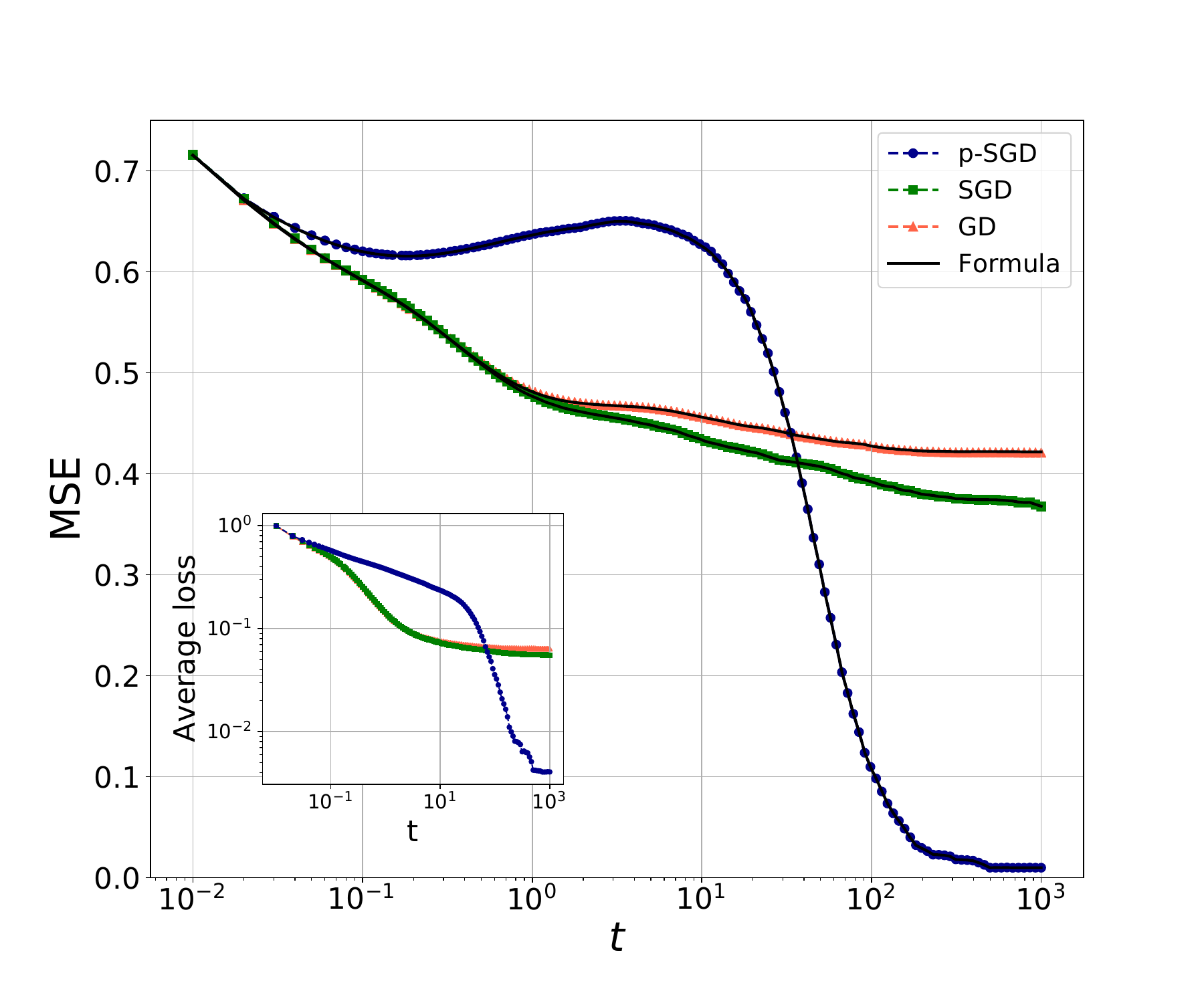}
\hspace{-7mm}
\includegraphics[scale=0.181]{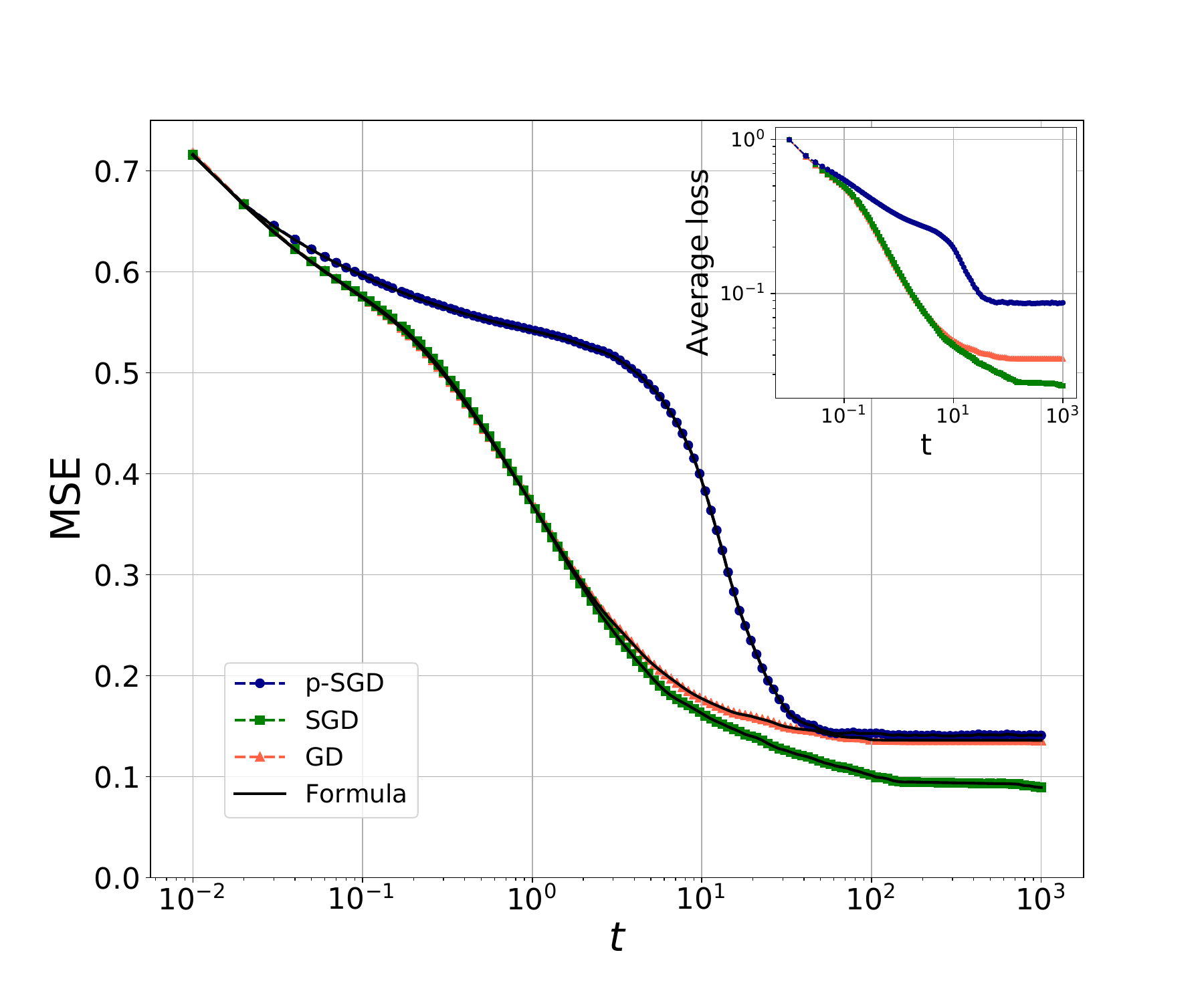}
\caption{\label{ridge_mse_varying_lambda} Mean squared error (main plot) and average training loss (inset) as a function of time, at dimension $N=1000$, $\alpha=M/N=3$, warm start $m_0=0.2$, learning rate $\eta=0.01$, for increasing values of regularization $\lambda=0$ (left), $\lambda=0.1$ (center) and $\lambda=1$ (right). We consider full-batch gradient descent (red triangles), multi-pass SGD (green squares) and its persistent version (blue dots). For the last two algorithms, we take a mini-batch size $\b=0.5$. The symbols mark the average over simulations, while the black line shows the prediction from Eq. \eqref{mse_formula}, where the overlaps $q_0$, $q$, and $m$ are obtained from simulations. The simulations are run over $250$ seeds and at each time a new landscape is generated. In both panels, the gradient is rescaled by the fraction of samples in the mini-batch at each iteration. }
\end{figure}

In this section, we sketch the computation of the average generalization error in the phase retrieval problem under consideration. Given a previously unseen data point $\underline{\xi}_{\rm new}\sim\mathcal{N}({\underline{0},\underline{\underline{I}}_N})$, the generalization error can be defined for a generic error function $f:\mathbb{R}^2\rightarrow \mathbb{R}$, taking as first argument the true label and as second argument the estimated one. The average generalization error is then:
\beq
\varepsilon_{\rm gen}=\mathbb{E}_{\{\underline{\xi}_\mu\}_{\mu=1}^{M},\underline{\xi}_{\rm new}, \underline{w}^{(0)}}\left[f(y_{\rm new},\hat{y}_{\rm new})\right],
\eeq 
where $y_{\rm new}=\biggr\rvert \frac{1}{\sqrt{N}}\underline{\xi}_{\rm new}\cdot\underline{w}^{(0)}\biggr\rvert$ is the true label, $\hat{y}_{\rm new}=\biggr\rvert \frac{1}{\sqrt{N}}\underline{\xi}_{\rm new}\cdot\underline{w}\biggr\rvert$ is the estimated one, and the weight vector $\underline{w}$ implicitly depends on the training set $\{\underline{\xi}_{\mu} \}_{\mu=1}^{M}$ as well as on the hidden signal $\underline{w}^{(0)}$.
We can introduce Dirac's $\delta-$functions to rewrite
\beq
\begin{split}
\varepsilon_{\rm gen}=&\mathbb{E}_{\{\underline{\xi}_\mu\}_{\mu=1}^{M},\underline{\xi}_{\rm new}, \underline{w}^{(0)}}\int_{-\infty}^{+\infty} \dde x \int_{-\infty}^{+\infty} \dde z \,f(|x|,|z|)\\&\times\delta\left(x-\frac{1}{\sqrt{N}}\underline{\xi}_{\rm new}\cdot\underline{w}^{(0)}\right)\delta\left(z-\frac{1}{\sqrt{N}}\underline{\xi}_{\rm new}\cdot\underline{w}\right)\\
=&\mathbb{E}_{\{\underline{\xi}_\mu\}_{\mu=1}^{M},\underline{\xi}_{\rm new}, \underline{w}^{(0)}}\int_{-\infty}^{+\infty} \frac{\dde x \dde \hat x}{2\pi} \int_{-\infty}^{+\infty}  \frac{\dde z \dde \hat z}{2\pi}  \,f(|x|,|z|)\\&\times\exp\left(i\hat x x +i\hat z z -\frac{i}{\sqrt N}\underline{\xi}_{\rm new}\cdot\left(\hat x \underline{w}^{(0)}+\hat z \underline{w}\right)\right),
\end{split}
\eeq
where we have substituted the $\delta-$functions with their Fourier representation.
We first compute the average over the new sample $\underline{\xi}_{\rm new}$, that is independent both of $\underline{w}^{(0)}$ and $\underline{w}$:
\beq
\begin{split}
\varepsilon_{\rm gen}=&\mathbb{E}_{\{\underline{\xi}_\mu\}_{\mu=1}^{M}, \underline{w}^{(0)}}\left[\int_{-\infty}^{+\infty} \frac{\dde x \dde \hat x}{2\pi} \int_{-\infty}^{+\infty}  \frac{\dde z \dde \hat z}{2\pi}  \,f(|x|,|z|)\right. \\ &\left. \times\exp\left(i\hat x x +i\hat z z -\frac{1}{2}\hat x ^2\frac{\underline{w}^{(0)}\cdot \underline{w}^{(0)}}{N} -\frac 12 \hat z ^2\frac{\underline{w}\cdot \underline{w}}{N}-\hat x \hat z \frac{\underline{w}^{(0)}\cdot \underline{w}}{N}\right)\right].
\end{split}
\eeq
In the following, we denote
\beq
q_0=\frac{\underline{w}^{(0)}\cdot \underline{w}^{(0)}}{N},\qquad
q=\frac{\underline{w}\cdot \underline{w}}{N},\qquad
m=\frac{\underline{w}^{(0)}\cdot \underline{w}}{N}.
\eeq
By integrating over the conjugate variables $\hat x$ and $\hat z$, we obtain
\beq 
\varepsilon_{\rm gen}=\mathbb{E}_{\{\underline{\xi}_\mu\}_{\mu=1}^{M}, \underline{w}^{(0)},z,x}\left[f(|x|,|z|)\right],\label{generalization_formula}
\eeq
where $z\sim\mathcal{N}(0,q)$ and $x\sim \mathcal{N}(my/q,q_0-m^2/q)$. In the infinite dimensional limit, $q_0$, $q$, and $m$ concentrate to their average value, therefore we simply obtain
\beq 
\varepsilon_{\rm gen}=\mathbb{E}_{z,x}\left[f(|x|,|z|)\right],
\eeq
where now the quantities $q_0,q,m$ are intended in the infinite dimensional limit. This computation shows that the generalization error depends on the signal and the training set only through $q_0,q,m$. In particular, in the spherical case $q_0=q=1$ and the performance depends only on $m$. 
\paragraph{Mean Squared Error} As a measure of the error, we can consider for instance the commonly-used mean squared error, here defined as:
\beq{\rm MSE}(y,\hat y)=\mathbb{E}(y-\hat y)^2.\eeq
From equation \eqref{generalization_formula}, we obtain that the mean squared error between the true label of a new sample and its estimate -- in the infinite dimensional limit -- is
\beq
{\rm MSE}=q+q_0-\frac 4\pi\left[\sqrt{qq_0-m^2}+m\,{\rm arctan}\left(\frac{m}{\sqrt{qq_0-m^2}}\right)\right],\label{mse_formula}
\eeq
which in the spherical case is a monotonically decreasing function of $m$. Fig. \ref{ridge_mse_varying_lambda} shows the mean squared error for gradient descent, multi-pass SGD and its persistent version in the case of ridge regularization. The dots mark the average from simulations, while the black line displays the prediction obtained from Eq. \eqref{mse_formula}, computed in the average values of $q_0$, $q$ and $m$ from simulations at dimension $N=1000$. We find a very good agreement between theory and simulations.

\begin{figure}
\centering
\includegraphics[scale=0.266]{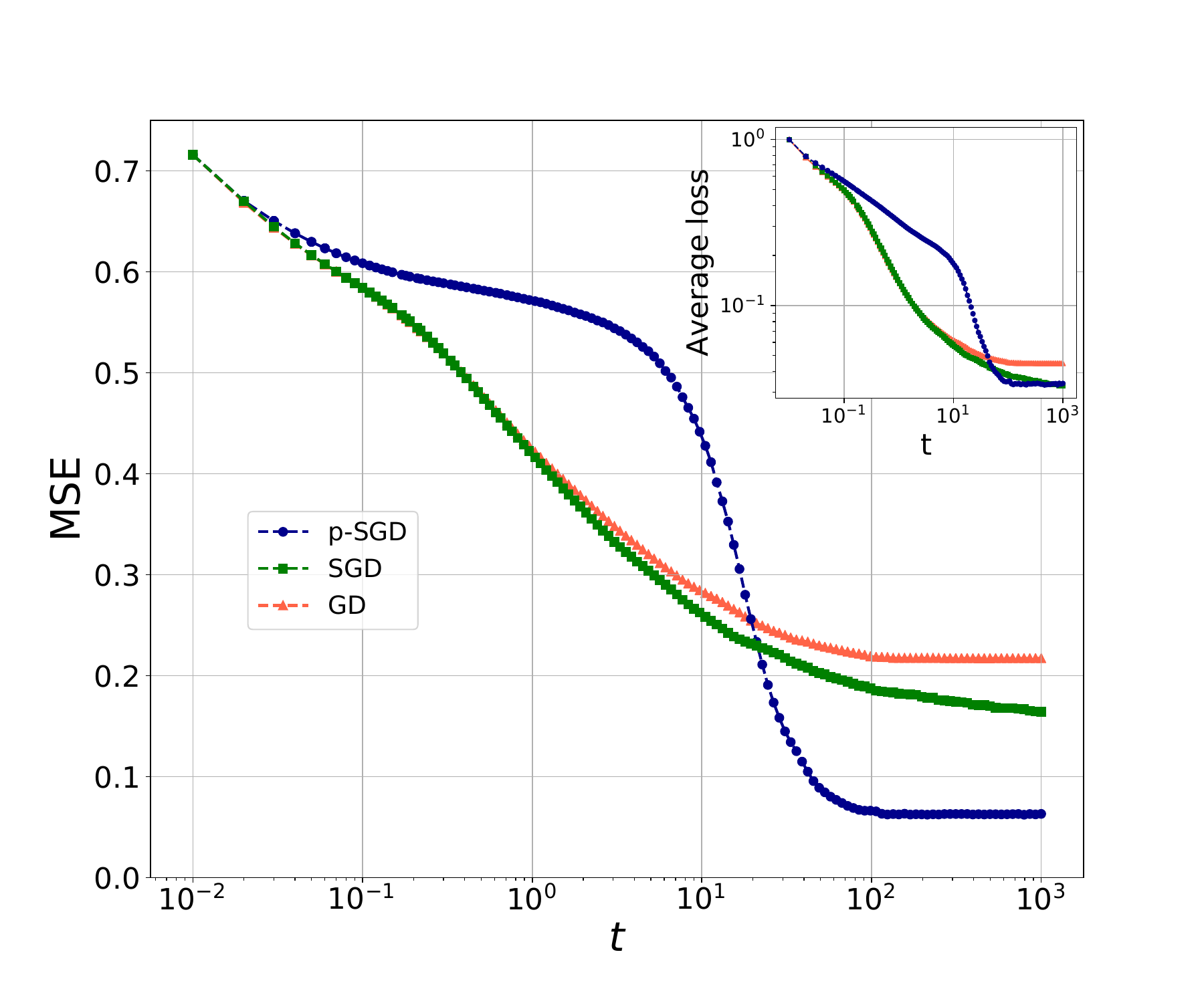}
\hspace{-8mm}
\includegraphics[scale=0.266]{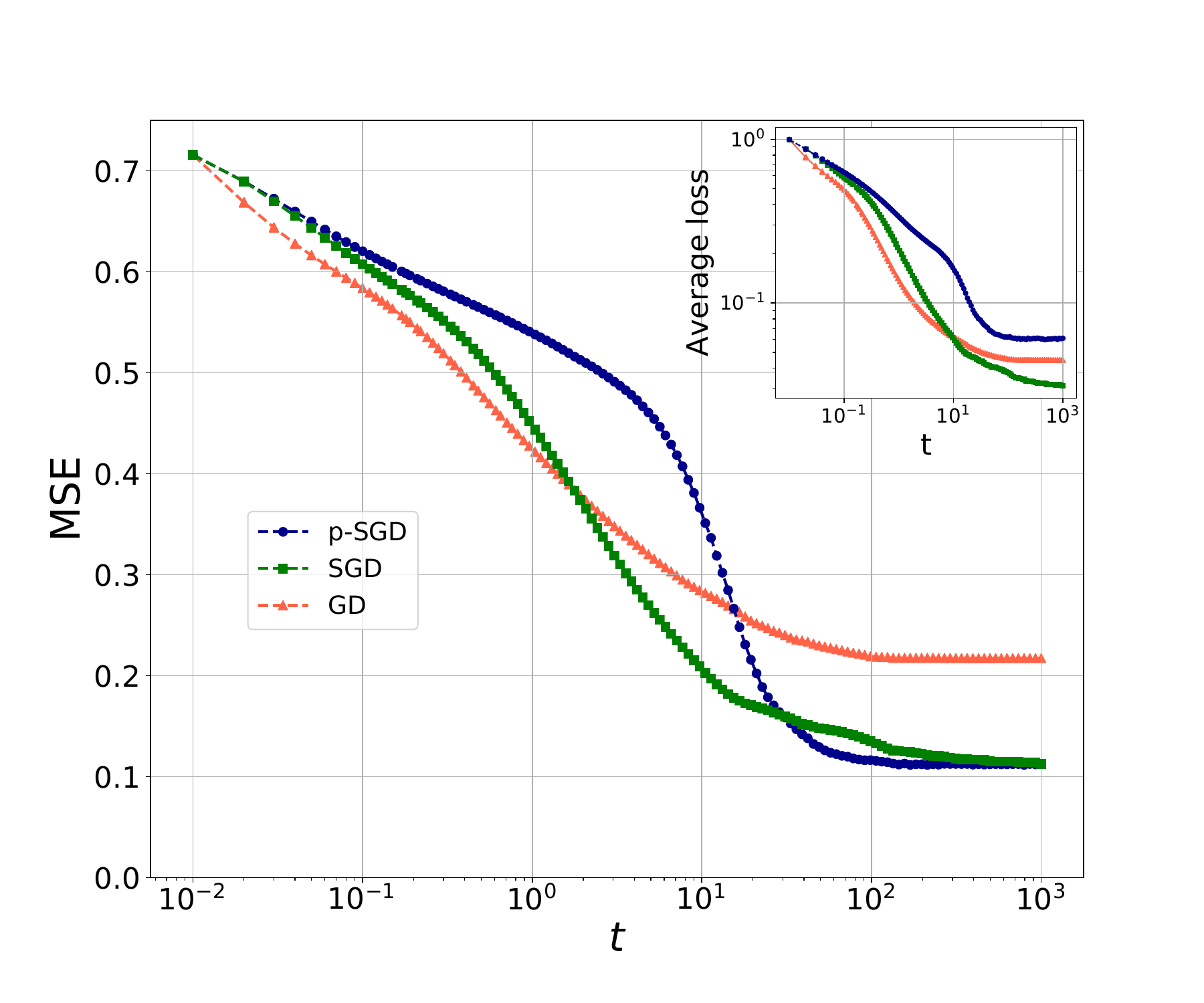}
\caption{\label{ridge_mse} Mean squared error (main plot) and average training loss (inset) as a function of time, at dimension $N=1000$, $\alpha=M/N=3$, warm start $m_0=0.2$, learning rate $\eta=0.01$, fixed regularization $\lambda=0.5$. We consider full-batch gradient descent (red triangles), multi-pass SGD (green squares) and its persistent version (blue dots). For the last two algorithms, we take a mini-batch size $\b=0.5$. The simulations are run over $250$ seeds and at each time a new landscape is generated. The left panel depicts the performance of the three algorithms when the gradient is rescaled by the fraction of training samples in the mini-batch $\b$, as defined in Eq. \eqref{flow_ridge}. The right panel illustrates the case in which the gradient is not rescaled by the fraction of samples used to compute it.  }
\end{figure}
\begin{figure}[!ht]
\begin{center}
\includegraphics[scale=0.266]{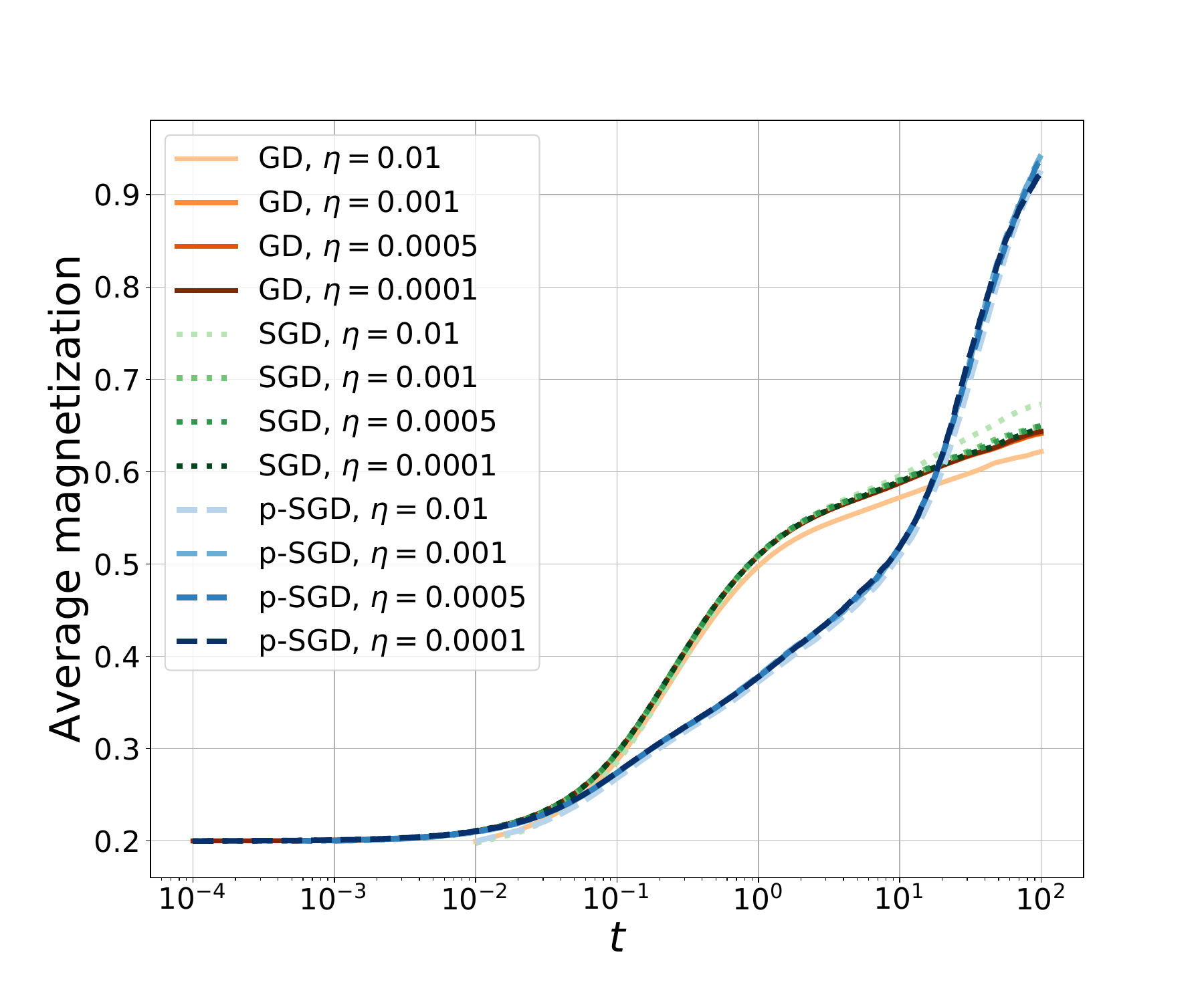}
\hspace{-10mm}
\includegraphics[scale=0.266]{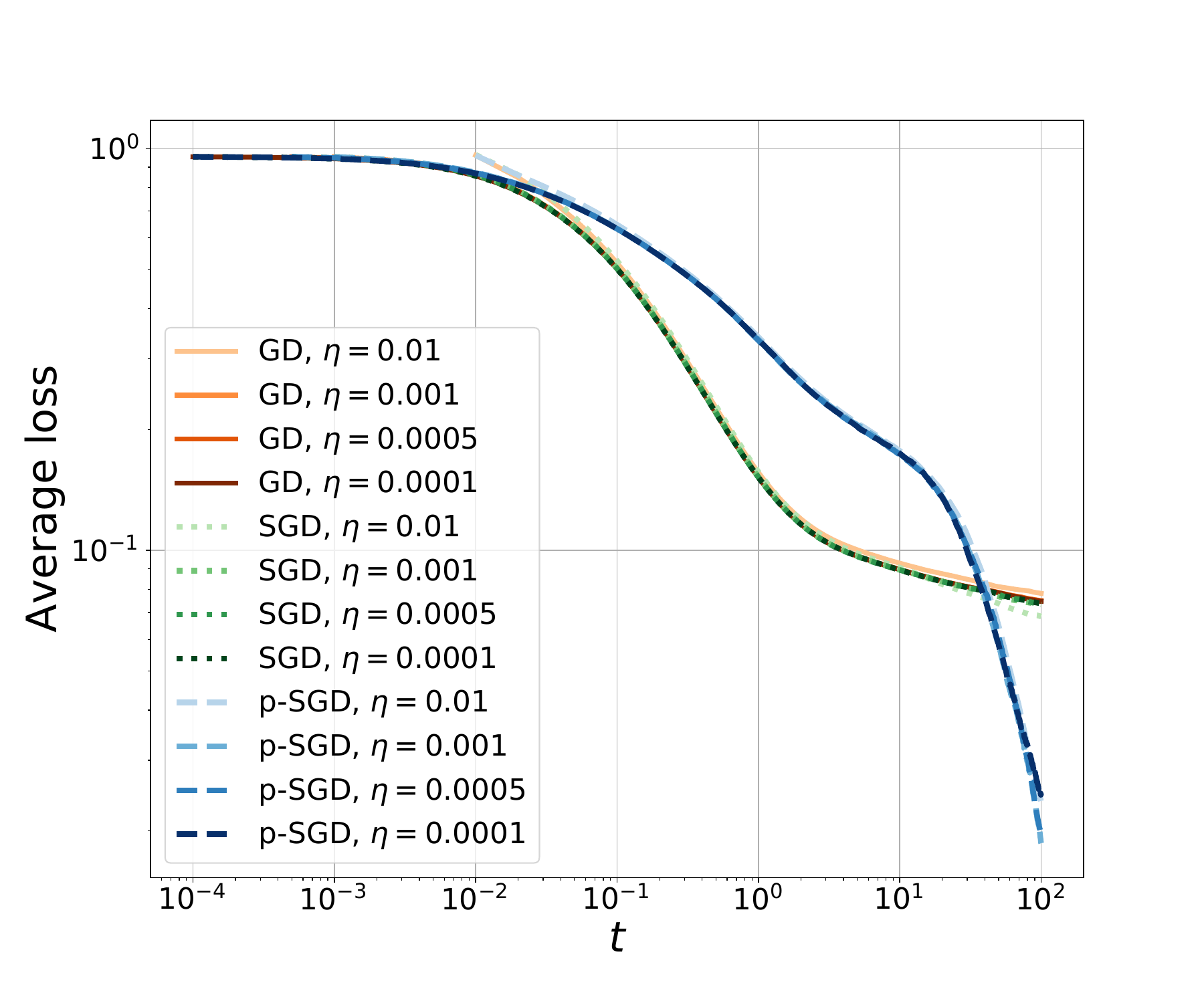}
\end{center}
\caption{\label{fig:small_lr}Average magnetization (right) and average loss (less) as a function of training time for the three algorithms: GD (full red lines),  SGD (dotted green lines) and persistent-SGD (dashed blue lines).  The numerical simulations are run at fixed $\alpha=M/N=3$, warm start $m_0=0.2$ and input dimension $N=1000$, over $250$ seeds. The stochastic algorithms are run at fixed batch size $\b=0.5$. We consider decreasing values of learning rate $\eta=0.01,0.001,0.0005,0.0001$, depicted with increasing color intensity. For visibility purposes, we plot $t+\eta$ on the x-axes.}
\end{figure}
\begin{figure}[!ht]
\begin{center}
\includegraphics[scale=0.266]{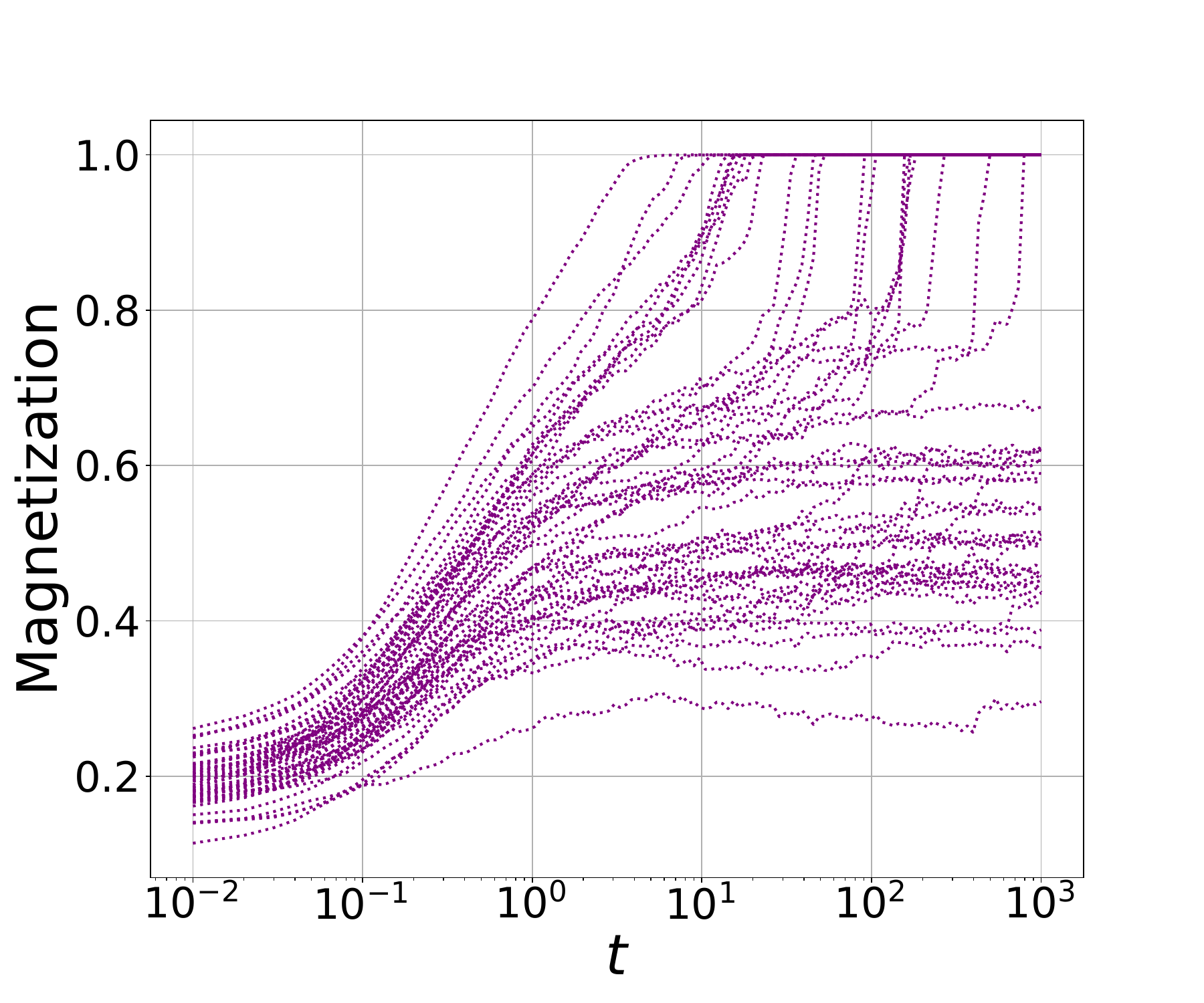}
\hspace{-10mm}
\includegraphics[scale=0.266]{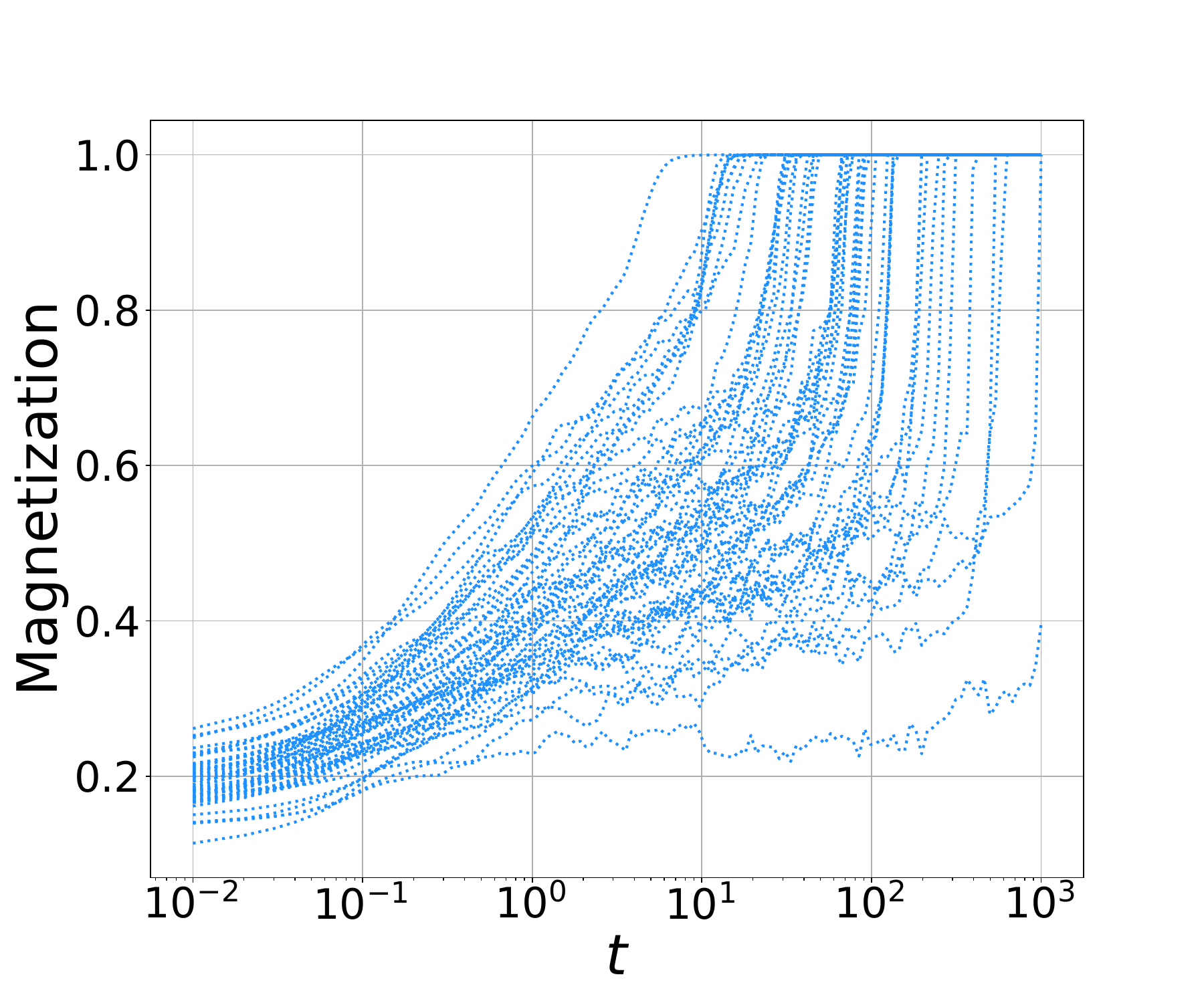}
\end{center}
\begin{center}
\includegraphics[scale=0.266]{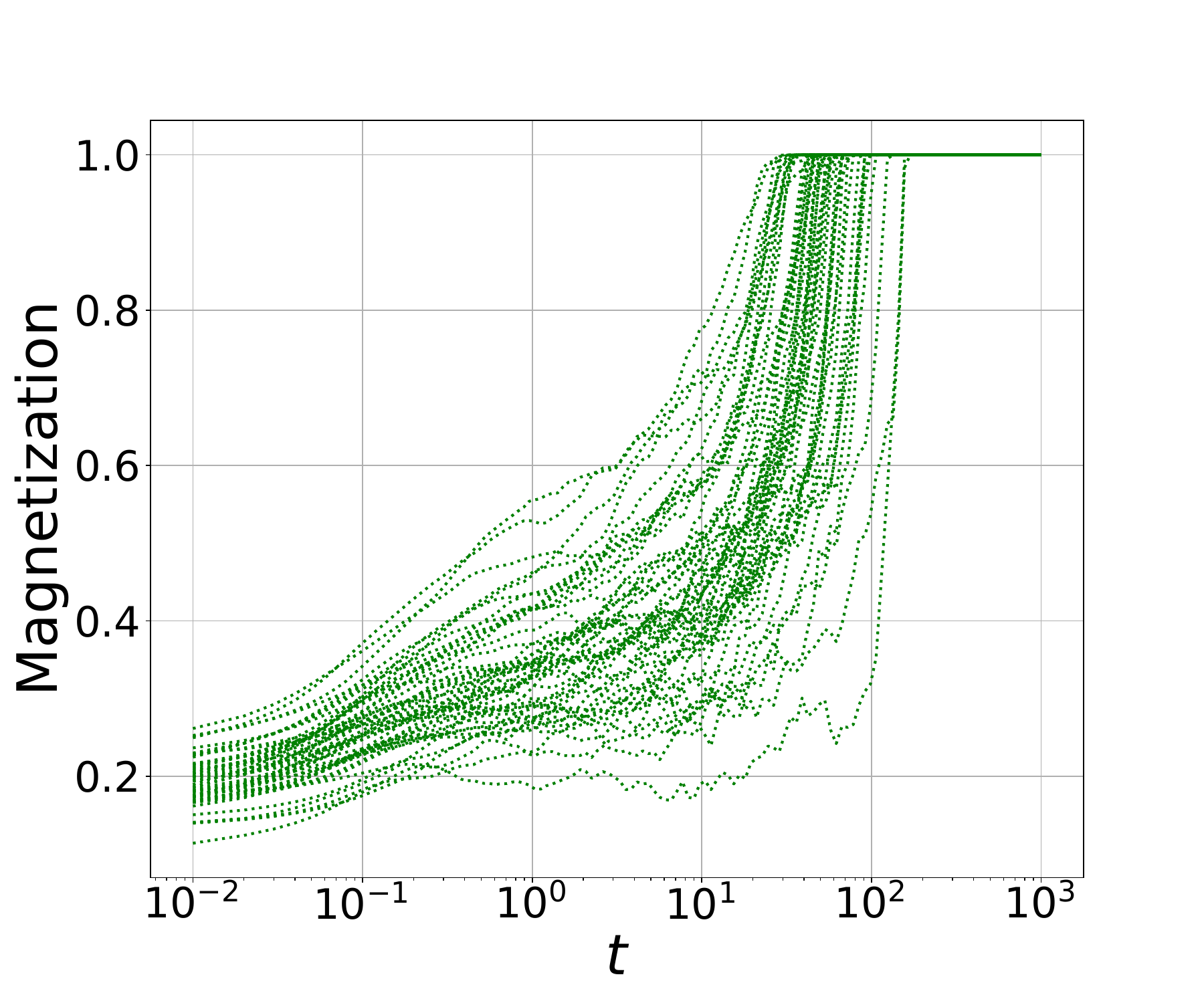}
\hspace{-10mm}
\includegraphics[scale=0.266]{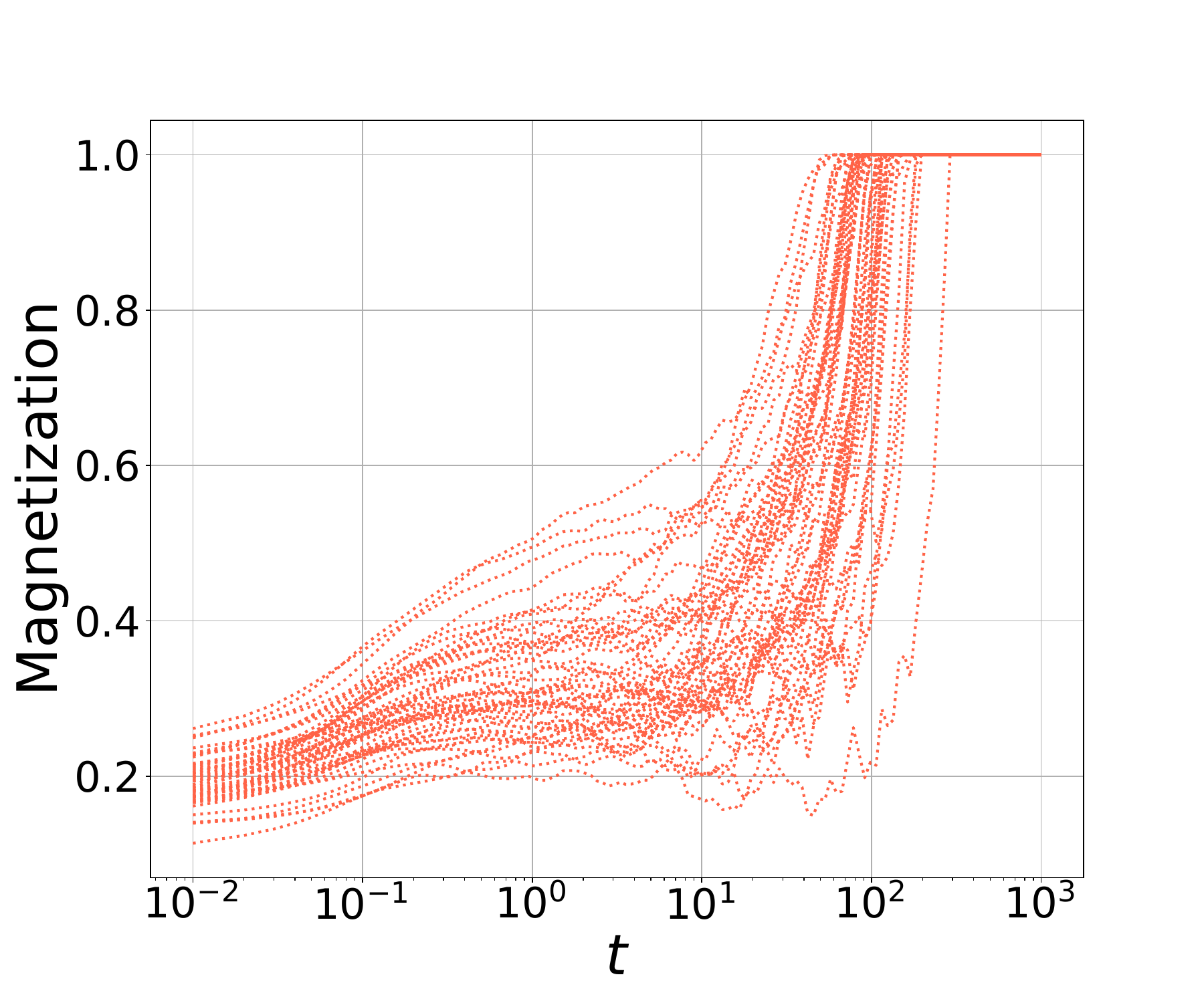}
\end{center}
\caption{\label{persistent_different_tau_instances} Instances of the magnetization as a function of time from numerical simulations for the persistent SGD algorithm at fixed $\alpha=M/N=3$, batch size $\b=0.5$ and warm initialization $m_0=0.2$. We consider the model with spherical constraint defined in Sec. \ref{sec:model_main} of the main text. We consider four different values of the persistence time: $\tau=0.05$ (upper left), $\tau=0.5$ (upper right), $\tau=2$ (lower left), $\tau=5$ (lower right). For each panel, we show $50$ different seeds, corresponding to different realizations of the landscape and initial weights. The simulations are run at dimension $N=1000$ and learning rate $\eta=0.01$.}
\end{figure}
\section{Ridge regularization}
\label{sec:ridge}
In this section, we consider a variant of the training algorithm presented in Eqs. \eqref{GDalgorithm}, \eqref{SGF}, and \eqref{Langevin} of the main text, where instead of projecting the weights on the hyper sphere $|\underline{w}(t)|^2=N$ at each iteration, we apply a ridge regularization of strength $\lambda$. The parameter $\lambda$ is fixed during training and can be tuned a posteriori, e.g., by cross-validation. The flow dynamics given by Eq. \eqref{continuous_equation} of the main text is modified as follows:
\beq
\frac{\partial w_i(t)}{\partial t} 
=-\lambda w_i(t) + \varsigma_i (t)-\frac 1\b\sum_{\mu=1}^{\a N} s_{\mu}(t)\partial_1 v(h_\mu(t);h_\mu^{(0)} )\frac{1}{\sqrt N} \xi^\mu_i.\label{flow_ridge}
\eeq
Note that this change simply amounts to substituting the time-dependent Lagrange multiplier $\hat \nu(t)$ with the constant $\lambda$. All the other variables are defined in the main text and stay the same. We modify accordingly the initial condition (defined by Eq. \eqref{init} of the main text) as follows:
\beq 
\underline{w}(t=0)=m_0\, \underline{w}^{(0)}+\underline{z} \in \mathbb{R}^N,
\eeq
where $m_0$ is the average initial magnetization and $\underline{z}\sim\mathcal{N}(\underline{0},\underline{I}_N)$. This change is reflected in the initial condition for the effective stochastic process in Eq. \eqref{supmat_eff_process}, that becomes : $P(h(0)) = e^{-h(0)^2/2}/\sqrt{2\pi }$. We still consider a teacher on the hyper sphere $|\underline{w}^{(0)}|^2=N$. 

\subsection{Results}
In this section, we discuss the results obtained for ridge
regularization. The behavior of the gradient-descent-based algorithms
is qualitatively the same as what we have observed for the spherical
case in the main text: stochasticity is beneficial for generalization
also in the case of ridge regularization and without any
regularization.

Fig. \ref{ridge_mse_varying_lambda} illustrates the performance of
gradient descent, multi-pass SGD and persistent SGD for three
different values of regularization strength: $\lambda=0$ (left panel),
$\lambda=0.1$ (central panel), $\lambda=1$ (right panel), fixing the
values of all the other control parameters. The generalization
performance is evaluated by measuring the MSE and the average training loss is
shown in the inset. We observe that the effect of ridge regularization
is different on SGD and persistent-SGD: while the former benefits from
a finite regularization $\lambda=1$, the latter generalizes better at
low values of $\lambda$. Evaluating the optimal regularization is
beyond the scope of this work. Furthermore, the left and central
panels of Fig. \ref{ridge_mse_varying_lambda} display a peculiar
phenomenon of \emph{double descent} of the generalization error as a
function of time that has also been observed in real data
\cite{Nakkiran2020Deep}.
 
Fig. \ref{ridge_mse} depicts the MSE as a function of time for the three algorithms under consideration at fixed regularization ($\lambda=0.5$) and for two different dynamics. In particular, we illustrate the effect of rescaling the gradient by the fraction of samples in the mini batch ($\b$) on the dynamics. In the left panel, the gradient is rescaled by $\b$, while in the right panel we do not rescale it. We observe that, while the rescaling is beneficial for persistent-SGD, SGD performs better without it. At variance with the spherical case considered in the main text, in the case of ridge regularization of fixed strength $\lambda$ rescaling the gradient by $\b$ does not result in a simple rescaling of the learning rate. Instead, the regularization is also affected.

\section{Additional figures}
\label{sec:numerical_simulations_supmat}
In this section we provide additional figures in support to our
observations in Sec. \ref{sec:analyzed_algorithms} and Sec. \ref{results_dynamics_main} of the main text. All
the figures illustrate the spherical case treated in the main
text. Therefore, the generalization performance is entirely captured
by the magnetization.

Fig. \ref{fig:small_lr} compares the average magnetization (left panel) and loss (right panel) as a function of training time for gradient-descent, SGD and persistent-SGD for decreasing values of the learning rate. We observe that, in the limit of small learning rate, the learning curves of SGD collapse to the ones of gradient descent. On the contrary,  the persistent-SGD algorithm has a well-defined continuous time limit that is different than the one of full batch gradient descent.

Fig. \ref{persistent_different_tau_instances} summarizes the effect of
increasing the persistence time on the performance of the
persistent-SGD algorihm. We show the instances of the magnetization as
a function of time -- corresponding to $50$ different realizations of
the problem landscape and initializations of the weight vector. We
consider increasing values of the parameter $\tau=0.05$ (upper left
panel), $\tau=0.5$ (upper right panel), $\tau=2$ (lower left panel),
and $\tau=5$ (lower right panel), at a fixed ratio $\alpha=3$ of
training samples over input dimensions, batch size $\b=0.5$ and warm
initialization $m_0=0.2$. On the one hand, we observe that increasing
the persistence time gradually diminishes the number of seeds that get
stuck at intermediate plateau, resulting in an improved generalization
performance. On the other hand, until time $t\sim \tau$ the samples in
the mini-batch have not been reshuffled yet (on average). Therefore,
for large values of $\tau$ the plateaus disappear but the
magnetization is stuck at the beginning of the training and only at
training time $t>\tau$ it has a sudden increase.

\end{document}